\documentclass[twoside]{bourlarge}
\usepackage{votrenom_h}
\usepackage{amssymb}
\usepackage{amsmath}
\usepackage{graphicx}

\newcommand{\be}{\begin{equation}}
\newcommand{\ee}{\end{equation}}

\newcommand{\ep}{\epsilon}
\def \a{\vec a}
\def \A{\vec A}
\def \K{\vec K}
\def \M{\vec M}
\def \KK{{\bf K}}
\def \DD{{\bf D}}
\def \MM{{\bf M}}
\def \G{\vec G}
\def \F{\vec F}
 \def \v{\vec v}
\def \D{\vec D}
\def \q{\vec q}
\renewcommand{\k }{\vec k}
\def \R{\vec R}
\def \r{\vec r}
\def \q{\vec q}
\def \p{\vec p}
\def \u{\vec u}
\def \sgn{{\rm sgn}}
	\newcommand{\bsq}{{\boldsymbol q}}

\begin{document}
\Large

\title{Dirac Fermions in condensed matter and beyond}

\author{Mark {\sc Goerbig} and Gilles {\sc Montambaux}\\
  Laboratoire de Physique des Solides -- B\^at. 510 \\ Universit\'e Paris Sud,
CNRS UMR 8502\\ 91405 Orsay cedex, France }

\maketitle

\begin{abstract}

This review aims at a theoretical discussion of Dirac points in two-dimensional systems. Whereas
Dirac points and Dirac fermions are prominent low-energy electrons in graphene (two-dimensional
graphite), research on Dirac fermions in low-energy physics has spread beyond condensed-matter
systems. In these alternative systems, a large versatility in the manipulation of the relevant
band parameters can be achieved. This allows for a systematic study of the motion and different
possible fusions of Dirac points, which are beyond the physical limits of graphene. We introduce
the basic properties of Dirac fermions and the motion of Dirac points here and aim at a
topological classification of these motions. The theoretical concepts are illustrated in
particular model systems.

\end{abstract}


\section{Introduction}

During the last decade, graphene research has triggered a tremendous interest in the physics
of two-dimensional (2D)
Dirac fermions in condensed matter physics \cite{grapheneRev}. Indeed, in undoped
graphene the valence band touches the conduction band at the Fermi level isotropically
in a linear manner. The low-energy band structure as well as the form of the underlying Hamiltonian
is reminiscent of that for massless fermions, usually studied in high-energy physics, with
two relevant differences. First, graphene electrons are constrained to move in two spatial dimensions,
whereas the framework of relativistic quantum mechanics was established to describe fermions in
three spatial dimensions. And second, the characteristic velocity that appears in condensed-matter
physics is not the speed of light but the Fermi velocity, which is roughly two orders of magnitude
smaller than the former. In addition to graphene, Dirac fermions have now been identified in various
other systems, both in condensed matter physics, such as at the surfaces of three-dimensional
topological insulators \cite{CarpentierIHP} or in quasi-2D organic materials \cite{ET3},
or in specially designed systems ("artificial graphenes"), such as e.g. cold atoms in optical lattices
\cite{Tarruell:12}, molecular crystals \cite{Manoharan:12} or microwave crystals \cite{Bellec:13a}.
Even if they are probably less promising for technological applications than graphene,
these artificial graphenes have the advantage that the relevant parameters can be varied more
easily. This versatility is the main motivation of this, mainly theoretical, review on Dirac
fermions in 2D systems, where we discuss the different manners of moving Dirac points in
reciprocal space and where we aim at a classification of the different types of Dirac-point merging.
Whereas we illustrate the theoretical concepts and this classification in several systems
realised experimentally, we do not aim at a complete account of artificial graphenes and
physical systems investigated in this framework.

The review is organised as follows. In Sec. \ref{sec:Emerg}, we discuss the general framework of
two-band Hamiltonians that may display Dirac points and investigate the role of discrete symmetries,
such as time-reversal and inversion symmetry. These considerations are the basis for an analysis
of the underlying (two-component) tight-binding models that describe Dirac fermions
(Sec. \ref{sec:TB}). In this section, we discuss both the topological properties of the spinorial
wave functions in the vicinity of Dirac points and the specific case of graphene. Section
\ref{sec:Bfield} is devoted to the behaviour of Dirac fermions in a strong magnetic field and the
relation between the topological winding properties of the wave functions and protected zero-energy
levels. In Sec. \ref{sec:MMDP}, we discuss the motion and merging of Dirac points related by
time-reversal symmetry and experimental implementations in cold atoms and microwave crystals. We
terminate this review with a more general discussion of how to obtain several pairs of Dirac
points in tight-binding models and a second class of Dirac-point merging that is topologically
different from that of Dirac points related by time-reversal symmetry (Sec. \ref{sec:NDP}).

\section{Emergence of Dirac fermions in a generic two-band model}
\label{sec:Emerg}

In lattice models, Dirac fermions
emerge at isolated points in the first Brillouin zone (BZ), where an upper band touches the lower one. The physically most interesting situation arises when the Fermi level precisely
resides in these contact points, as for example in undoped graphene \cite{grapheneRev}. On quite
general grounds, a two-band model that could reveal Dirac points may be expressed in terms of the
band Hamiltonian

\be {\cal{H}}_{\k}= \sum_{\mu=0}^3 f_{\k}^{\mu}\sigma^{\mu}=\left(
                \begin{array}{cc}
                  f_{\k}^0+f_{\k}^{z} & f_{\k}^x-i f_{\k}^y \\
                  f_{\k}^x+if_{\k}^y & f_{\k}^0 - f_{\k}^z \\
                \end{array}
              \right) \label{GenHam}  \ee
in reciprocal space, where $\sigma^0$ is the $2\times 2$ one matrix and
\be
\nonumber
\sigma^x=\left(\begin{array}{cc} 0 & 1\\ 1 & 0 \end{array}\right), \qquad
\sigma^y=\left(\begin{array}{cc} 0 & -i\\ i & 0 \end{array}\right),\qquad
\sigma^z=\left(\begin{array}{cc} 1 & 0\\ 0 & -1 \end{array}\right)
\ee
are Pauli matrices. Because the band Hamiltonian must be a Hermitian matrix, the functions
$f_{\k}^{\mu}$ are real functions of the wave vector
that reflect furthermore the periodicity of the underlying lattice.
The two bands are easily obtained from a diagonalisation of Hamiltonian (\ref{GenHam}) that yields
\be\label{eq:GenBands}
\tilde{\epsilon}_{\lambda}(\k)=f_{\k}^0 +\lambda\sqrt{(f_{\k}^x)^2 + (f_{\k}^y)^2 + (f_{\k}^z)^2},
\ee
where $\lambda=\pm 1$ is the band index. One notices that the function $f_{\k}^0$ is only an
offset in energy, and we define for convenience the band energy
$\epsilon_{\lambda,\k}=\tilde{\epsilon}_{\lambda,\k}-f_{\k}^0$ -- since the function $f_{\k}^0$
in the Hamiltonian goes along with the one matrix, it does not affect the (spinorial) eigenstates
obtained from a diagonalisation of the Hamiltonian. One clearly sees from the generic expression
(\ref{eq:GenBands}) for the two bands, that band contact points require the annihilation of all
three functions $f_{\k}^x$, $f_{\k}^y$ and $f_{\k}^z$ at isolated values $\k_D$ of the wave vector.
In a 2D space, one is thus confronted with a system of three equations
\be
\nonumber
f_{\k_D}^x=0, \qquad f_{\k_D}^y=0, \qquad f_{\k_D}^z=0
\ee
to determine two values, i.e. the components of the wave vector $\k_D=(k_D^x,k_D^y)$. In contrast
to three spatial dimensions, where the equations would determine three parameters and where
one would then obtain three-dimensional Weyl fermions \cite{WeylRev}, stable band contact points
can only be obtained in 2D when one of the components is equal to zero over a larger interval
of wave vectors. This situation arises when particular symmetries are imposed on the system that
we discuss in the next paragraph.

One of the symmetries that protect stable band-contact points in Hamiltonian (\ref{GenHam}) is
time-reversal symmetry, and it imposes
\be\label{eq:TR}
{\cal{H}}_{-\k}^*={\cal{H}}_{\k}\qquad \rightarrow\qquad f_{-\k}^x=f_{\k}^x,
\qquad f_{-\k}^y=-f_{\k}^y, \qquad f_{-\k}^z=f_{\k}^z.
\ee
Notice that in this argument, we have omitted the spin degree of freedom. Since we consider here a system
with no spin-orbit coupling,  each of the bands (\ref{eq:GenBands}) is simply two-fold
degenerate.\footnote{Otherwise, time-reversal symmetry would read
${\cal{H}}_{-\sigma,-\k}^*={\cal{H}}_{\sigma,\k}$ where $\sigma$ denotes the orientation of the
electronic spin, and the band Hamiltonian is thus necessarily a $4\times 4$ matrix. A detailed
discussion of this case is beyond the scope of the present review.}
Another relevant symmetry is inversion symmetry. Consider that the diagonal elements of Hamiltonian
(\ref{GenHam}) represent intra-sublattice (or intra-orbital) couplings. The first spinor component
would then correspond to the weight on the $A$ sublattice and the second one to that on the $B$
sublattice. Inversion symmetry imposes that the Hamiltonian be invariant under the exchange of the
two sublattices $A\leftrightarrow B$ while interchanging $\k\leftrightarrow -\k$,
in which case one finds the conditions
\be\label{eq:IR}
f_{-\k}^x=f_{\k}^x, \qquad f_{-\k}^y=-f_{\k}^y, \qquad f_{-\k}^z=-f_{\k}^z.
\ee
for the periodic functions. Whereas the first two conditions are compatible with those (\ref{eq:TR})
obtained for time-reversal symmetry, the presence of both symmetries (inversion and time-reversal)
yields a function $f_{\k}^z$ that is both even and odd in the wave vector, i.e. it must
eventually vanish for all wave vectors, $f_{\k}^z=0$. In the remainder of this review, we will
therefore restrict the discussion to the Hamiltonian

\be {\cal{H}}_{\k}= \left(
                \begin{array}{cc}
                  0 & f_{\k} \\
                  f^*_{\k} & 0 \\
                \end{array}
              \right), \label{GH}  \ee
which respects both symmetries and where we have defined the complex function
$f_{\k}=f_{\k}^x-if_{\k}^y$.

\section{Dirac fermions in tight-binding models and fermion doubling}
\label{sec:TB}

Before discussing the specific lattice model relevant for graphene, let us consider some general
aspects of Hamiltonian (\ref{GH}) in the framework of general tight-binding
parameters. Indeed the non-zero components of the band Hamiltonian reflect the
periodicity of the Bravais lattice and may be written quite
generally in the form

\be f_{\k}= \sum_{m,n} t_{mn} e^{- i {\vec k} \cdot \R_{mn}} \
, \label{fofk} \ee
where the hopping amplitudes
$t_{mn}$ are real, a consequence of the time-reversal
symmetry in Eq. (\ref{eq:TR}), and   $\R_{mn}=
m \a_1 + n \a_2$ are vectors of the underlying Bravais lattice.
The Dirac points, which we coin $\DD$ and $\DD'$ are
solutions of the complex equation $f_{\D}=0$ (see above).
Since $f_{\k}=f^*_{-\k}$, the
Dirac points, when they exist, necessarily come in by pairs, i.e.
if $\D$ is a solution, so is $-\D$. Whereas this is a natural situation in
lattice systems and gives rise to a $2N$-fold valley degeneracy (in the case of $N$
pairs of Dirac points), this happens to be a problematic situation in high-energy
physics, where Dirac fermions (of continuous systems) are sometimes simulated in lattice models
and where fermions are thus doubled artificially \cite{fermiondoubling}.
The positions  of the Dirac points can be
anywhere in the BZ and move upon variation of the band parameters
$t_{mn}$. Around the Dirac points $\pm \D$, the function $f_{\k}$ varies linearly.
Writing ${\vec k} = \pm \D+\q$, we find, in a system of units with $\hbar=1$ that we
adopt henceforth,

\be f_{\pm \D+\q}=\q \cdot (\pm {\vec v}_1 - i {\vec v}_2)
 \label{fofq}
\ee
where the velocities ${\vec v}_1$ and ${\vec v}_2$ are given by
\be
{\vec v}_1  =  \sum_{mn} t_{mn} \R_{mn} \sin \D \cdot \R_{mn} \qquad {\rm and} \qquad
{\vec v}_2  =  \sum_{mn} t_{mn} \R_{mn} \cos \D \cdot \R_{mn}
\label{v1v2}
\ee
Furthermore, one generally has ${\vec v}_1  \neq {\vec v}_2$ so that the Dirac cones are not necessarily isotropic, and the low-energy
Hamiltonian in the vicinity of the Dirac points $\xi\D$ is written
as
\be\label{eq:DirHam}
{\cal{H}}_{\q}^{\xi}=\q\cdot\left(\xi{\vec v}_1\sigma^x +{\vec v}_2\sigma^y\right),
\ee
where $\xi$ is the valley index ($\xi=+$ for $\DD$ and $\xi=-$ for $\DD'$). Naturally, if there
are other pairs of Dirac points, one obtains pairs of Hamiltonians of type (\ref{eq:DirHam})
for each of them. The above analysis will further help us in the discussion of Dirac-point motion
and merging presented in Sec. \ref{sec:MMDP}. In the vicinity of the Dirac points, the dispersion
relation is then given by
\be
\epsilon_{\lambda}(\q)=\lambda\sqrt{(\vec{v}_1\cdot\q)^2+(\vec{v}_2\cdot\q)^2}.
\ee
Notice that the case $\vec{v}_1\parallel\vec{v}_2$ is pathological in the sense that there
would be no dispersion in the direction perpendicular to $\vec{v}_1$ and $\vec{v}_2$. This would
be a quasi-1D limit that we exclude in the following discussions.\footnote{It would require an
expansion of $f_{\k}$ beyond linear order around the Dirac points
to obtain a dispersion in this direction.}

\subsection{Rotation to a simplified model and spinorial form of the wave functions}

The Dirac Hamiltonian (\ref{eq:DirHam}) may be further simplified and brought to the form
\be\label{eq:DirHamB}
{\cal{H}}_{\q'}^{\xi}=\xi v_x'q_x'\sigma^{x\prime}+v_y'q_y'\sigma^{y\prime}
\ee
with the help of a rotation of the coordinate space
\begin{eqnarray}
\nonumber
q_x &=& \cos\vartheta q_x' + \sin\vartheta q_y'\\
\nonumber
q_y &=& -\sin\vartheta q_x' +\cos\vartheta q_y',
\end{eqnarray}
accompanied by a rotation of the (sublattice)-pseudospin frame around the $z$-quantisation axis
\begin{eqnarray}
\nonumber
\sigma^x &=& \cos\theta \sigma^{x\prime} + \sin\theta \sigma^{y\prime}\\
\nonumber
\sigma^y &=& -\sin\theta \sigma^{x\prime} +\cos\theta \sigma^{y\prime},
\end{eqnarray}
and in terms of the novel velocities \cite{ET3}
\begin{eqnarray}
\nonumber
v_x^{\prime 2} &=& \frac{|\vec{v}_1|^2 + |\vec{v}_2|^2}{2}
+\frac{1}{2} \sqrt{|\v_1|^4 + |\v_2|^4 + 2(\v_1\cdot\v_2)^2 - 2(\v_1\wedge\v_2)^2
}
\ ,\\
\nonumber
v_y^{\prime 2} &=& \frac{|\vec{v}_1|^2 + |\vec{v}_2|^2}{2}
-\frac{1}{2} \sqrt{|\v_1|^4 + |\v_2|^4 + 2(\v_1\cdot\v_2)^2 - 2(\v_1\wedge\v_2)^2}
\ .
\end{eqnarray}
Here, $\vec{v}_1\wedge\vec{v}_2=(\vec{v}_1\times\vec{v}_2)_z$ is the $z$-component of the
3D vector product $\vec{v}_1\times\vec{v}_2$.
In the remainder of this section, we omit the primes at the velocities and wave vectors
assuming that we are in the appropriate frame after transformation.

The eigenenergies of the low-energy model (\ref{eq:DirHamB}) are simply
\be
\epsilon_{\lambda}(\q)=\lambda\sqrt{v_x^2q_x^2 + v_y^2q_y^2},
\ee
with the corresponding eigenstates
\be\label{eq:b0states}
\psi_{\xi,\lambda;\q}=\frac{1}{\sqrt{2}}\left(
\begin{array}{c}
1 \\ \xi\lambda e^{-i\phi_{\q}},
\end{array}\right)
\ee
where the relative phase between the two components depends on the orientation of the wave
vector,
\be\label{eq:RelPh}
\tan\phi_{\q}=\frac{v_yq_y}{v_xq_x}.
\ee

\subsection{Berry phases and winding numbers}
\label{sect:winding}

The relative phase $\phi_{\q}$ derived above in Eq. (\ref{eq:RelPh}) exhibits  a particular topological structure encoded in the Berry phase \cite{berry}. Around each Dirac point, the circulation of  $\phi_{\q}$ along a closed path is quantized; the quantity
\be
w_{\xi,\lambda}= {\xi\lambda\sgn(v_xv_y) \over 2 \pi} \oint_{C_i}  \nabla_{\q} \phi_{\q} \cdot d\k
\ee
is an integer, the topological winding number associated with each Dirac point.\footnote{
This quantitiy is, modulo a factor of $\pi$, nothing other than the Berry phase accumulated on
the path $C_i$ around the Dirac point. However, it is more convenient to use the concept of
(topological) winding numbers -- whereas one is used, in basic quantum mechanics, to the fact that
a phase $2\pi$ is identical to 0, the winding number is a physically relevant quantity
(as we show below) that makes a clear distinction between $w_i=0$ and $2$.} With the help
of Eq. (\ref{eq:RelPh}), one finds that the winding number associated with a
constant-energy path at either positive ($\lambda=+1$) or negative ($\lambda=-1$) energy
around the Dirac point $\xi$ is simply
\be
w_{\xi,\lambda}=\xi\lambda\sgn(v_xv_y).
\ee
An important observation is the fact that, because
the two Dirac $\DD$ and $\DD'$ are related
by time-reversal symmetry, they have \textit{opposite} winding numbers.

This argument may be generalised to situations with several pairs of Dirac points related by
time-reversal symmetry. Consider $N$ pairs of Dirac points
(situated at the positions $\xi\D_i$ in the first Brillouin zone, $i=1,..,N$),
each of which is described at low energy
by a Hamiltonian of the type (\ref{eq:DirHam}),
\be
{\cal{H}}^{i,\xi}_{\q}=\q\cdot\left(\xi{\vec v}_1^i\sigma^x +{\vec v}_2^i\sigma^y\right).
\ee
Applying the same arguments around these novel points yields a winding number
\be
w_{\xi,\lambda}^i=\xi\lambda\frac{\vec{v}_1\wedge\vec{v}_2}{|\vec{v}_1\wedge\vec{v}_2|}.
\ee

The winding number, which is a conserved quantity,
may be interepreted as a topological charge. It is
additive, and we will therefore use this concept extensively when discussing the different
types of Dirac-point motion and merging in Sec. \ref{sec:MMDP}. Most saliently, it provides a
simple and convenient manner of identifying the number of topologically protected Dirac points and
zero-energy states, namely in a magnetic field, as we will show below.

\subsection{Basic properties of electrons in graphene}

\begin{figure}[htbp!]
\begin{center}
\includegraphics[width=5cm]{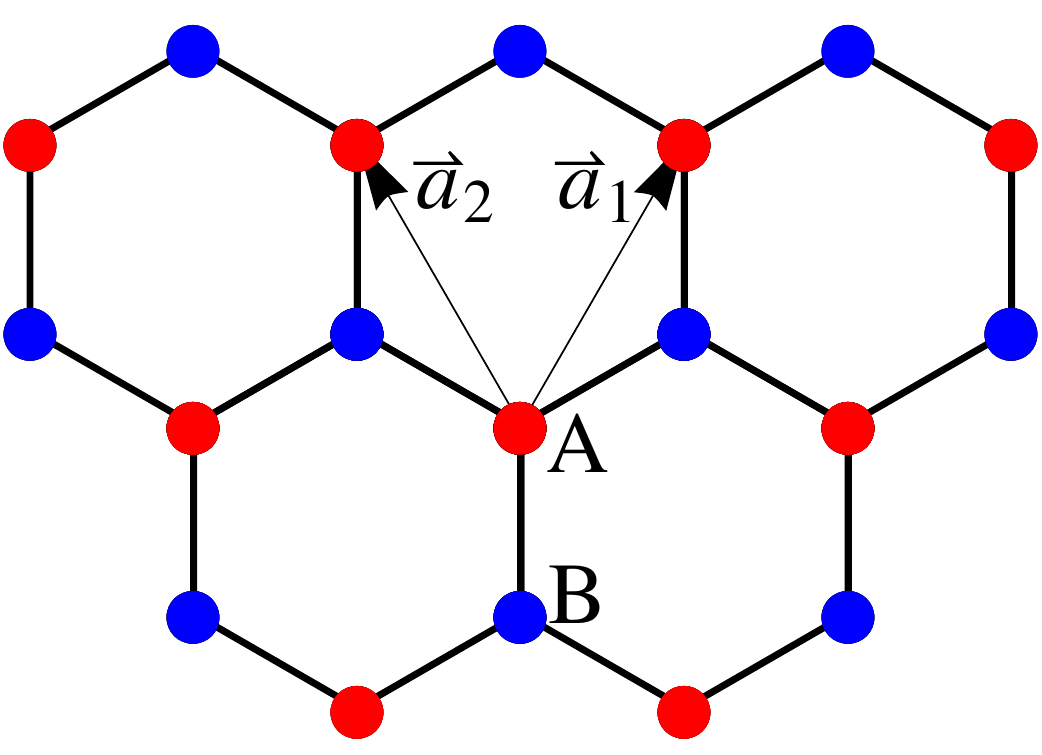}\hspace*{0.2cm}
\includegraphics[width=8cm]{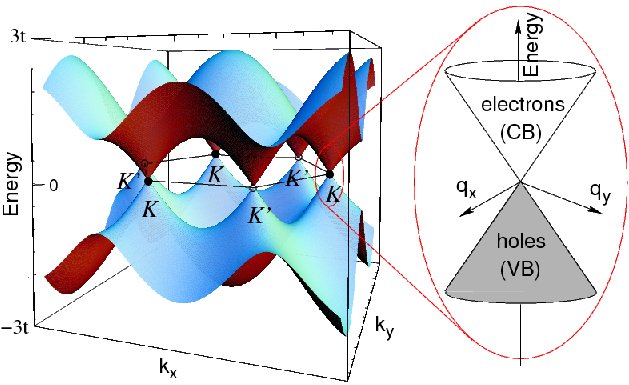}
\end{center}
\caption{Left: Honeycomb lattice of graphene. The simplified tight-binding model takes into account
hopping between nearest-neighbour sites, e.g. from A to B$_1$, B$_2$ and $B_3$. In the case of
undistorted graphene, the associated three hopping amplitudes are identical due to the point
symmetry of the lattice. Right: Energy bands of graphene obtained from the tight-binding model
and zoom around the Dirac point at $K$.}
\label{fig:honeycomb1}
\end{figure}

Before discussing these zero-energy states in a magnetic field, let us use the above considerations
to analyse the band structure of graphene \cite{wallace},
the probably best known instance of 2D Dirac points in
condensed matter. Graphene consists of a one-atom thick layer of carbon atoms
arranged in a honeycomb lattice (see left pannel of Fig. \ref{fig:honeycomb1} for a sketch of the
lattice structure).
The underlying Bravais lattice is thus a triangular lattice, and the honeycomb structure is obtained
with the help of a two-atom basis (sites $A$ and $B$). The low-energy electronic properties
of graphene can be obtained to great accuracy within a simplified tight-binding model, where hopping
only between nearest-neighbour $p_z$ orbitals
is taken into account, with a characteristic energy scale of
$t\simeq 3$ eV. The band Hamiltonian (\ref{GenHam}) thus only has off-diagonal terms, and the function
(\ref{fofk}) simply reads
\be
f_{\k}=t\left(1 + e^{i\k\cdot\a_1} + e^{i\k\cdot\a_2}\right),
\ee
where $\a_1= (\sqrt{3}a/2)(\u_x + \sqrt{3}\u_y)$ and $\a_2=(\sqrt{3}a/2)(-\u_x + \sqrt{3}\u_y) $ are basis vectors that span the triangular Bravais lattice (see Fig.
\ref{fig:honeycomb1}). Here, $\u_x$ and $\u_y$ are the unit vectors in the $x$- and $y$- direction,
respectively, and $a\simeq 0.14$ nm is the distance between nearest-neighbour carbon atoms.
Notice that, because both sublattices $A$ and $B$ consist of carbon atoms, inversion symmetry is
respected as well as time-reversal symmetry. Corrective tight-binding terms that are neglected in the
present model therefore do not generate a non-zero term $f_{\k}^z$.\footnote{Next-nearest neighbour
hopping breaks particle-hole symmetry by generating a term $f_{\k}^0$ but does not open a gap.}
The band structure of graphene is depicted in Fig. \ref{fig:honeycomb1} (right pannel),
and one notices the characteristic Dirac
points at the corners of the first Brillouin zone,
\be
\K=\frac{4\pi}{3\sqrt{3}a}\u_x \qquad {\rm and} \qquad \K'=-\K=- \frac{4\pi}{3\sqrt{3}a}\u_x,
\ee
where the Fermi level resides in the absence of doping.\footnote{Indeed, the electronically
relevant $p_z$ orbitals, which give rise to the electronic bands, are each
occupied by a single electron such that the band structure is half-filled, and the Fermi level is
thus situated at the Dirac points.}
Notice furthermore that there are four additional contact points in the dispersion relation visible
in Fig. \ref{fig:honeycomb1}.
However, these additional points are connected to $\K$ and $\K'$ by a reciprocal lattice
vector so that they correspond to the same electronic  state. The fact that the Dirac points
coincide with the corners of the Brillouin zone in the case of graphene is a consequence of the
crystal point symmetry -- as we will show in the following sections, deviations from this symmetry
place the Dirac points at less symmetric points of the BZ, but the Dirac points are
nevertheless topologically protected by inversion and time-reversal symmetry.

Similarly to the general case, the low-energy Hamiltonian is readily obtained by expanding
$f_{\k}$ as in Eq. (\ref{fofq}) around the Dirac points,
\be
f_{\pm \D+\q}=v_F(q_x-iq_y),
\ee
where $v_F=3at/2$ is the Fermi velocity. Compared to Eq. (\ref{fofq}), the point symmetry of the
undistorted graphene lattice provides us with isotropic Dirac points, $\v_1=(v_F,0)$ and
$\v_2=(0,v_F)$, and Hamiltonian (\ref{eq:DirHam}) becomes
\be\label{eq:DirHamGraph}
{\cal H}_{\q}^{\xi}=v_F(\xi q_x\sigma^x + q_y\sigma^y).
\ee
This Hamiltonian is precisely that of massless Dirac fermions of relativistic quantum mechanics in
two spatial dimensions. The description of low-energy electrons in graphene within this framework
has been extremely successful in identifying its original electronic properties. A full account
on these properties is beyond the scope of the present review, and we refer the reader to existing
reviews. Apart from the general ones \cite{grapheneRev,chakraRev}, there are reviews that concentrate
on more specific aspects of graphene, such as \cite{peresRev} and \cite{DSRev} for electronic
transport, \cite{expRev} and \cite{SiCRev} for experimental reviews, as well as \cite{kotovRev} and
\cite{goerbigRev} for interaction effects in graphene.

\section{Dirac fermions in a magnetic field}
\label{sec:Bfield}

Instead of providing an exhaustive account on physical phenomena of graphene electrons, we discuss
here the consequences of the above-mentioned topological aspects in the presence of a magnetic
field. Most saliently, the winding numbers allow us to identify the degeneracy of the zero-energy
states (zero-energy Landau levels) and their topological stability.

\subsection{Landau levels of Dirac fermions}

The basic model Hamiltonian (\ref{eq:DirHamB}), transformed to the appropriate reference frame
as described above, is amenable to an exact quantum-mechanical solution when taking into account
a perpendicular magnetic field via the Peierls substitution,
\be\label{eq:Peierls}
\q \rightarrow \vec{\Pi}=\p + e\A(\r),
\ee
where $\p$ is the quantum-mechanical momentum (in the continuum limit), which is conjugate to
a coarse-grained position operator $\r$. This Peierls substitution is valid as long as the
magnetic field $B=|\nabla\times\A(\r)|$ is sufficiently small, such that the associated magnetic
length $l_B=1/\sqrt{eB}\simeq 26\,{\rm nm}/\sqrt{B{\rm [T]}}$ is much larger than the
lattice spacing, $l_B\gg a$. This condition is ususally satisfied in condensed matter physics at
accessible magnetic fields (up to 45 T for static magnetic fields and $B\lesssim 200$ T for
pulsed fields in the semi-destructive regime). The quantum-mechanical commutation relations
$[x,p_x]=[y,p_y]=i$ (and 0 for the crossed commutators) induce the commutation relations
\be
[\Pi_x,\Pi_y]=-\frac{i}{l_B^2}
\ee
for the components of the gauge-invariant kinetic momementum. In terms of the associated ladder
operators
\be\label{eq:ladder}
\hat{a}=\frac{l_B}{\sqrt{2}}(\Pi_x - i\Pi_y) \qquad {\rm and}
\qquad \hat{a}^{\dagger}=\frac{l_B}{\sqrt{2}}(\Pi_x + i\Pi_y),
\ee
which satisfy $[\hat{a},\hat{a}^{\dagger}]=1$, the Peierls substitution can be simplified to
\be
q \rightarrow \frac{\sqrt{2}}{l_B}\hat{a} \qquad {\rm and} \qquad q^*\rightarrow
\frac{\sqrt{2}}{l_B}\hat{a}^{\dagger},
\ee
and the Hamiltonian (\ref{eq:DirHamB}) can eventually be written as
\be\label{eq:DirB}
H_B^{\xi=+}=\sqrt{2}\frac{v_F}{l_B}\left(
\begin{array}{cc}
0 & \hat{a} \\ \hat{a}^{\dagger} & 0
\end{array}
\right) \qquad {\rm and} \qquad
H_B^{\xi=-}=-\sqrt{2}\frac{v_F}{l_B}\left(
\begin{array}{cc}
0 & \hat{a}^{\dagger} \\ \hat{a} & 0
\end{array}
\right) ,
\ee
where the Fermi velocity $v_F$ is an average quantity $v_F=\sqrt{v_xv_y}$.

The energy spectrum of the so-called Landau levels can be obtained directly with the help
of the eigenstates of the number operator $\hat{n}=\hat{a}^{\dagger}\hat{a}$,
\be\label{eq:LLs}
\epsilon_{\lambda,n}=\lambda \frac{v_F}{l_B}\sqrt{2n},
\ee
and reveals the characteristic $\sqrt{Bn}$ scaling \cite{ML}
observed in magneto-spectroscopic measurements
\cite{grenoble} and scanning-tunneling spectroscopy \cite{LL_STS}. Notice that the spectrum does
not depend on the valley index $\xi$, and one obtains thus a two-fold valley degeneracy in addition
to the usual spin degeneracy (unless the latter is lifted by a strong Zeeman effect).
The corresponding eigenstates read
\be\label{eq:statesN}
\psi_{\lambda,n\neq 0}^{\xi=+}=\frac{1}{\sqrt{2}}\left(
\begin{array}{c}
|n-1,m\rangle \\ \lambda |n,m\rangle
\end{array}
\right) \qquad {\rm and} \qquad
\psi_{\lambda,n\neq 0}^{\xi=-}=\frac{1}{\sqrt{2}}\left(
\begin{array}{c}
|n,m\rangle \\ -\lambda |n-1,m\rangle
\end{array}
\right),
\ee
for the levels with $n\neq 0$.
Here, the spinor components satisfy $\hat{n}|n,m\rangle= n|n,m\rangle$, and an additional
quantum number $m$ needs to be taken into account to complete the basis. We discuss the physical
meaning of this quantum number below.
The zero-energy Landau levels $n=0$ need to be treated separately
and reveal a very special structure,
\be\label{eq:statesN0}
\psi_{n= 0}^{\xi=+}=\left(
\begin{array}{c}
0 \\ |n=0,m\rangle
\end{array}
\right) \qquad {\rm and} \qquad
\psi_{n= 0}^{\xi=-}=\left(
\begin{array}{c}
|n=0,m\rangle \\ 0
\end{array}
\right),
\ee
and one notices that, at zero energy, the valley degree of freedom coincides with the sublattice
index. At zero energy, the dynamical properties of the A sublattice (electrons in valley $K'$)
are thus completely decoupled from that on the B sublattice (electrons in valley $K$). This gives
rise to the particular series of quantum Hall effects at filling factors
\be\label{eq:FF}
\nu=\frac{n_{el}}{n_B}=2\pi l_B^2 n_{el}=\pm 2(2n +1),
\ee
where $n_{el}$ is the 2D electronic density and $n_{B}=1/2\pi l_B^2=eB/h$ the density of flux quanta
threading the graphene sheet. The effect manifests itself by plateaus in the transverse (Hall)
resistance, at magnetic fields corresponding to these filling factors, accompanied by zeros in
the longitudinal resistance. The observation of quantum Hall states at the filling factors
(\ref{eq:FF}) in 2005 \cite{novoselov,kim} was interpreted as a direct proof of the presence of
pseudo-relativistic carriers in graphene described in terms of massless Dirac fermions
(\ref{eq:DirHamGraph}).

\subsection{Degeneracy}

We have already alluded to the presence of a second quantum number $m$ in the full description
of the quantum-mechanical states (\ref{eq:statesN}) and (\ref{eq:statesN0}). Since the
Landau level spectrum (\ref{eq:LLs}) does not depend on this quantum number, its presence yields
a degeneracy of the Landau levels that we discuss in more detail here. Indeed,
the kinetic momentum $\vec{\Pi}$ introduced in the Peierls substitution (\ref{eq:Peierls}) may
be related to the cyclotron variable $\vec{\eta}$ with the components
\be
\eta_x=l_B^2\Pi_y \qquad {\rm and} \qquad \eta_y=-l_B^2\Pi_x,
\ee
that satisfy in turn the commutation relation $[\eta_x,\eta_y]=-il_B^2$. In classical mechanics,
this cyclotron variable describes precisely
the cyclotron motion of a charged particle in a uniform magnetic field. In addition, one knows from
classical mechanics that the particle's energy does not depend on the position of the centre of
the cyclotron motion, which is thus a constant of motion. This gauge-invariant
centre of the cyclotron motion,
\be\label{eq:Dec}
\R=(X,Y)=\r - \vec{\eta},
\ee
which is also called \textit{guiding centre}, remains a constant of motion in
the quantum-mechanical description, i.e. its components commute with the Hamiltonian
$[X,H_{B}^{\xi}]=[Y,H_{B}^{\xi}]=0$. Furthermore, one can show from the decomposition (\ref{eq:Dec})
that the components of the guiding centre commute with those of the cyclotron variable
\be
[\eta_x,X]=[\eta_y,X]=[\eta_x,Y]=[\eta_y,Y]=0,
\ee
whereas the guiding-centre coordinates do not commute among each other,
\be\label{eq:GC}
[X,Y]=il_B^2.
\ee
This allows for the introduction of a second set of ladder operators
\be
\hat{b}=\frac{1}{\sqrt{2}l_B}(X+iY) \qquad {\rm and} \qquad
\hat{b}^{\dagger}=\frac{1}{\sqrt{2}l_B}(X-iY),
\ee
with $[\hat{b},\hat{b}^{\dagger}]=1$, similarly to those (\ref{eq:ladder}) introduced in the
description of the kinetic momentum. The second quantum number $m$ is thus simply the eigenvalue
of $\hat{b}^{\dagger}\hat{b}$, and describes, as mentioned above, the orbital
degeneracy of the Landau levels. Instead of deriving explicitly this degeneracy,\footnote{Whereas
the general proof is rather involved, the degeneracy can be obtained when analysing
the wave functions in a particular gauge \cite{goerbigLH}.}
one may invoke an argument via the Heisenberg uncertainty relation associated with the
commutation relation (\ref{eq:GC}),
\be\label{eq:Heisenberg}
\Delta X\Delta Y \gtrsim 2\pi l_B^2=\sigma.
\ee
This means that each quantum-mechanical state $|n,m\rangle$ occupies a minimal surface $\sim\sigma$,
and the degeneracy of each Landau level may thus be quantified by dividing the full area
$\Sigma$ by this minimal surface,
\be\label{eq:LLdeg}
N_B=\frac{\Sigma}{\sigma}=n_B\Sigma,
\ee
in terms of the flux density $n_B=1/2\pi l_B^2=eB/h$, which we have already encountered in the previous
paragraphs. The filling factor (\ref{eq:FF}) can thus be interpreted as the number of Landau levels
that are completely filled, while not taking into account its internal degeneracy due to the spin and
valley degrees of freedom. The latter four-fold degeneracy indicates that there are eventually
$4N_B$ states per Landau level and the quantum-Hall plateaus thus occur at multiples of four of
the filling factor, as suggested by Eq. (\ref{eq:FF}).
Notice finally that, in principle, the Heisenberg uncertainty relation
(\ref{eq:Heisenberg}) is an inequality and that $\sigma$ would then just be a lower bound of the
surface occupied by a quantum state. However, the above-mentioned calculation in a special geometry and
a special gauge \cite{goerbigLH} indicates that the minimal surface is $\sigma$ and that the
degeneracy is indeed given by Eq. (\ref{eq:LLdeg}). As a qualitative explanation of this fact, one
may invoke the harmonic-oscillator structure of the quantum-mechanical system -- the associated
wave functions are therefore Gaussians for which the Heisenberg uncertainty relation becomes an
equality.

Notice finally that the arguments in this subsection rely only on the algebraic structure of
2D electrons in a magnetic field and not of the precise form of the Hamiltonian. As long as the
latter can be expressed solely in terms of the kinetic-momentum operator $(\Pi_x,\Pi_y)$, each
of its (perhaps unspecified) energy levels is $N_B$-fold degenerate, apart from internal degrees of
freedom, such as the spin, or a ``topological'' degeneracy that we will discuss in the following
subsection. This particular feature is simply due to the existence of a second set of
operators, $X$ and $Y$, that commute with $\Pi_x$ and $\Pi_y$ but that do not commute among each
other [Eq. (\ref{eq:GC})].

\subsection{Semi-classical quantisation rule}

In the previous subsections, we have discussed the Landau level spectrum of massless Dirac fermions
in monolayer graphene and shown that these levels are highly degenerate as a consequence of the
existence of a set of operators, $X$ and $Y$, that commute with the Hamiltonian. However, many
electronic systems do not have a simple low-energy description in terms of the 2D Dirac Hamiltonian
(\ref{eq:DirHam}) (or a simple Schr\"odinger equation) that allows for an exact quantum-mechanical
solution in the presence of a magnetic field. In this case, one needs to appeal to other methods,
such as the semi-classical quantisation rule \cite{Onsager,Degail:12}

\be
{\mathcal A}_{\mathcal C}(\ep_n) =\frac{2 \pi}{l_B^2}
\left( n + {1 \over 2} - {|w_{\mathcal C}| \over 2}\right) \ ,  \label{SCQR}
\ee
where ${\mathcal A}_{\mathcal C}(\ep_n)$ is the area in reciprocal space
delimited by a closed  contour $\mathcal{C}$, associated with energy $\ep_n$.
Furthermore, $w_{\mathcal C}$ represents the total winding of the relative phase $\phi_{\k}$
between the spinor components along this closed contour. In the case of massless Dirac
fermions, this phase was given by Eq. (\ref{eq:RelPh}), whereas in the general two-band
model (\ref{eq:DirHamB}) it reads
\be
\tan \phi_{\k}=\frac{{\rm Im} f_{\k}}{{\rm Re} f_{\k}},
\ee
in terms of the real and the imaginary parts of the complex function $f_{\k}$.

One can easily show that the calculation of the reciprocal-space area
${\mathcal A}_{\mathcal C}=2\pi \int_0^{k_n}dk\,k = 2\pi \int_0^{\ep_n}d\ep\,k(\ep)(dk/d\ep)$
yields the graphene Landau-level spectrum (\ref{eq:LLs}) for isotropic Dirac points,
with $\epsilon=v_F k$, for which we have already calculated the winding number,
$w=\pm 1$. Beyond the calculation of the Landau-level spectrum, the semi-classical analysis
(\ref{SCQR}) is also extremely convenient in the classification of the different types of
Dirac point motion and merging discussed in the following section. Indeed, the winding numbers
are topological charges and thus conserved quantities in the different merging scenarios. The
merging of two Dirac points with opposite winding numbers therefore gives rise to a total
winding of $w=0$, that is zero topological charge, and that of Dirac points with like winding yield
a total topological charge of $|w|=2$. We will discuss both types of merging transitions extensively
in the following sections and finish this paragraph with another consequence of the topological charge.
Indeed, it classifies the number of topologically protected zero-energy levels, which is
\be
w_{p}=\left|\sum_i w_i\right|,
\ee
in the case of $i=1,...,2N$ Dirac points in the system. As we have already mentioned, if the system
is time-reversal symmetric (in the absence of a magnetic field), this number is necessarily zero
because Dirac points emerge in pairs of opposite charge. However, if the Dirac points are sufficiently
isolated at low magnetic fields, there are
\be
w_t=\sum_i |w_i|
\ee
(i.e. $w_t=2N$) zero-energy levels associated with the $N$ pairs of Dirac Hamiltonians of the type
(\ref{eq:DirB}) that describe the low-energy electronic excitations of the system.

\section{Motion and merging of time-reversal-symmetric Dirac points}
\label{sec:MMDP}

\begin{figure}[htbp!]
\begin{center}
\includegraphics[width=4cm]{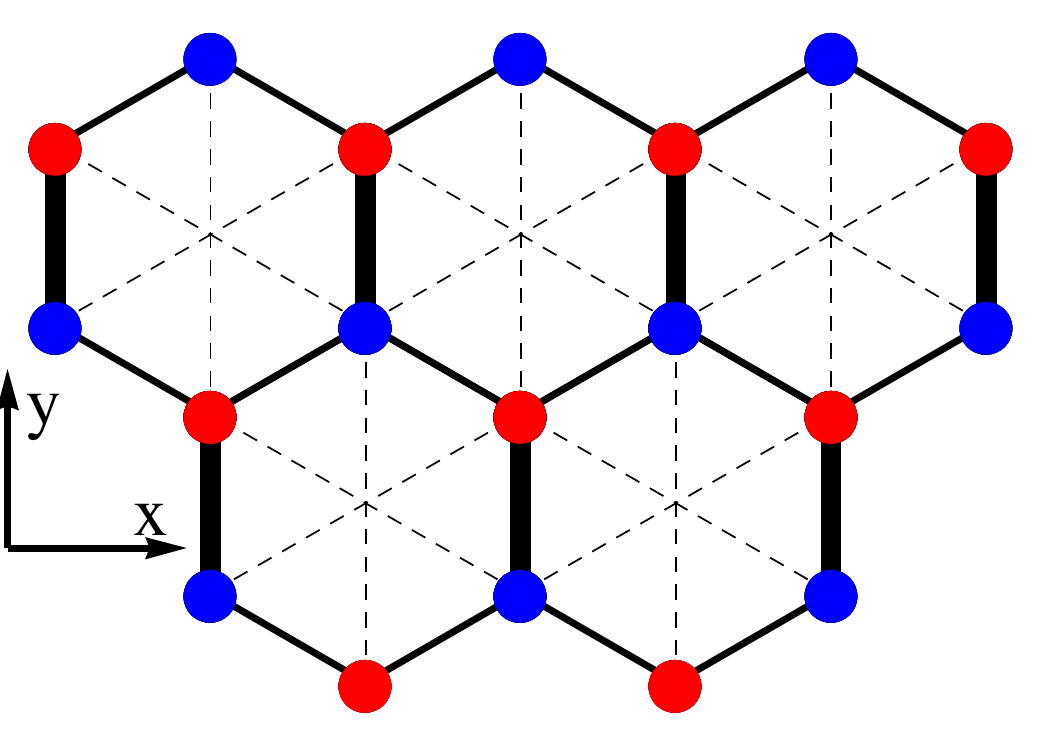}
\includegraphics[width=3.5 cm]{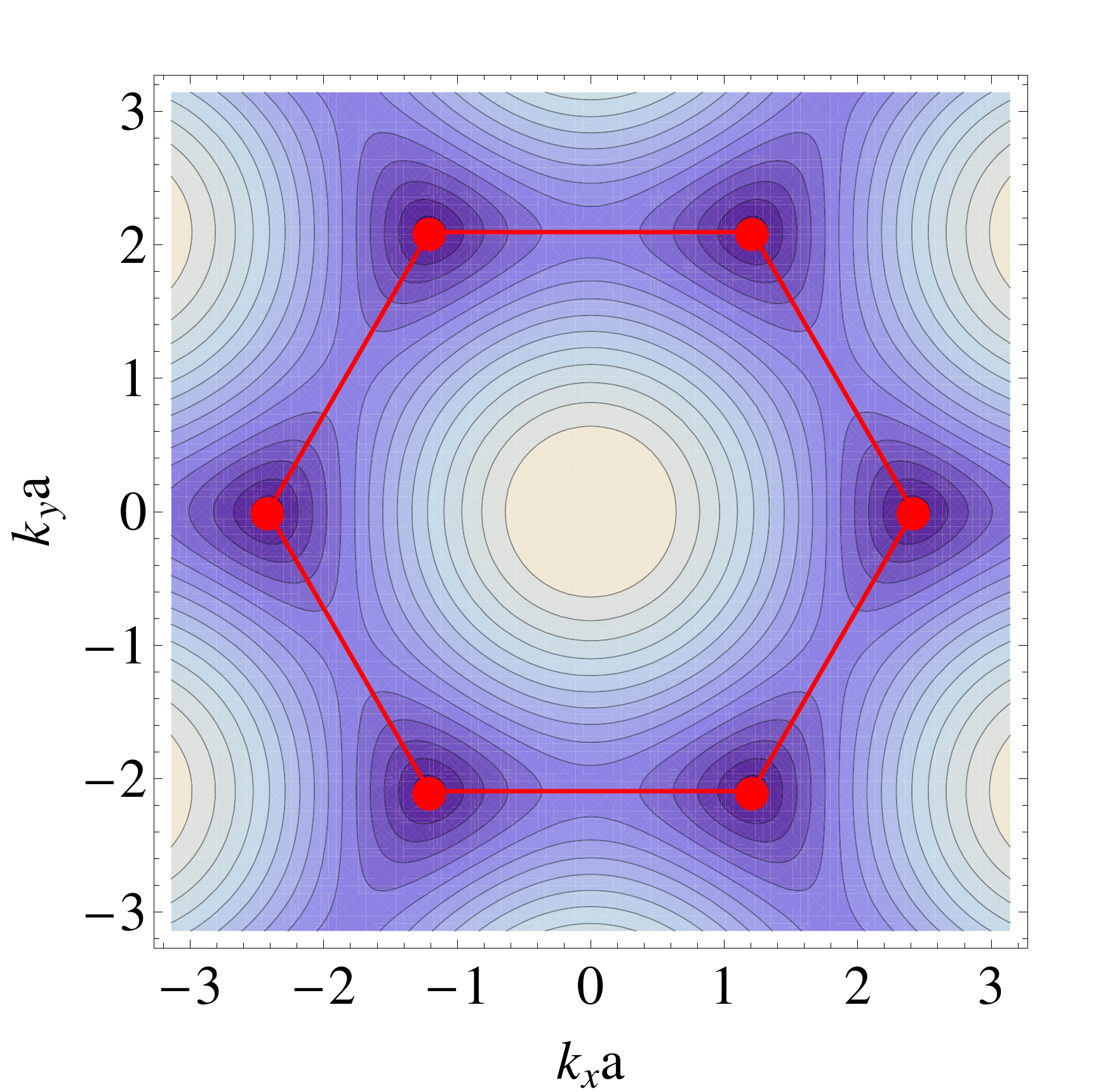}\includegraphics[width=3.5 cm]{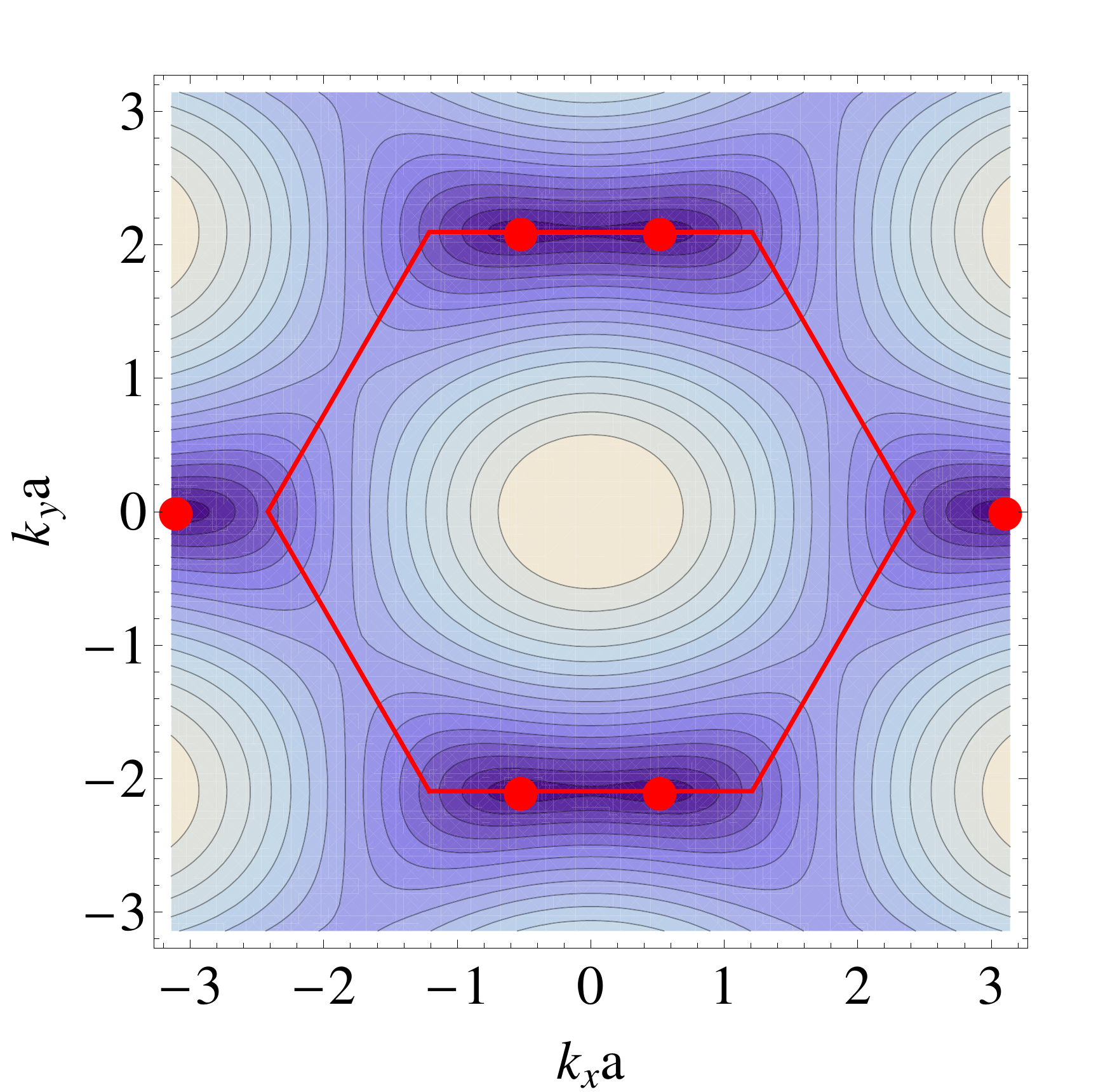}\includegraphics[width=3.5cm]{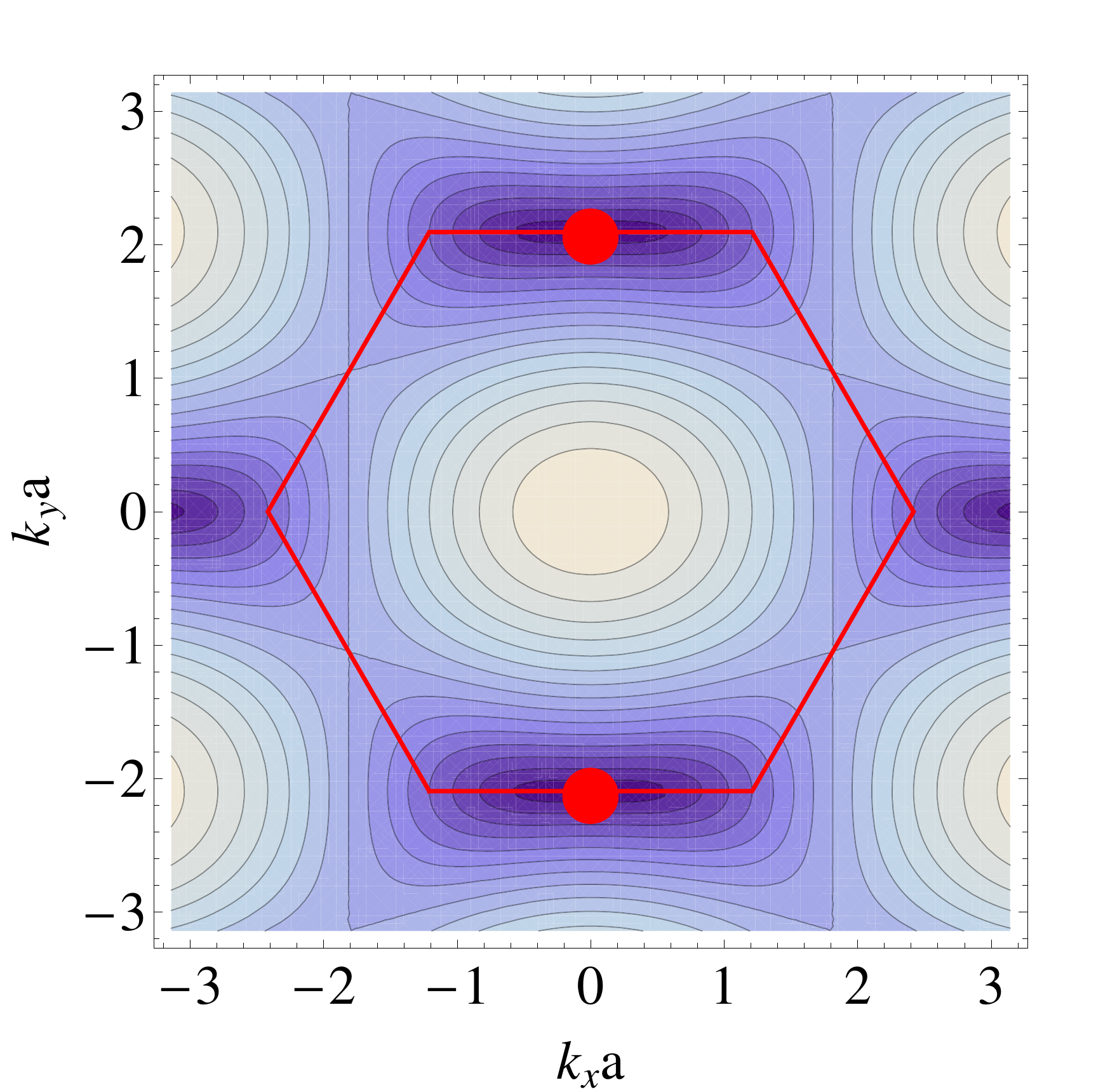}
\end{center}
\caption{Left~: honeycomb crystal. The vertical thick lines represent the modified hopping integral $t'$. The dashed lines indicate the third nearest neighbour coupling $t_3$ discussed in section \ref{sect:tt3}. Right~: variation of the isoenergy lines for three values of $t'/t=1, 1.8, 2$ ($t_3=0$). The red dots indicate the position of the two inequivalent Dirac points which merge when $t'=2t$.}
\label{fig:honeycomb}
\end{figure}

\begin{figure}[htbp!]
\begin{center}
\includegraphics[width=4cm]{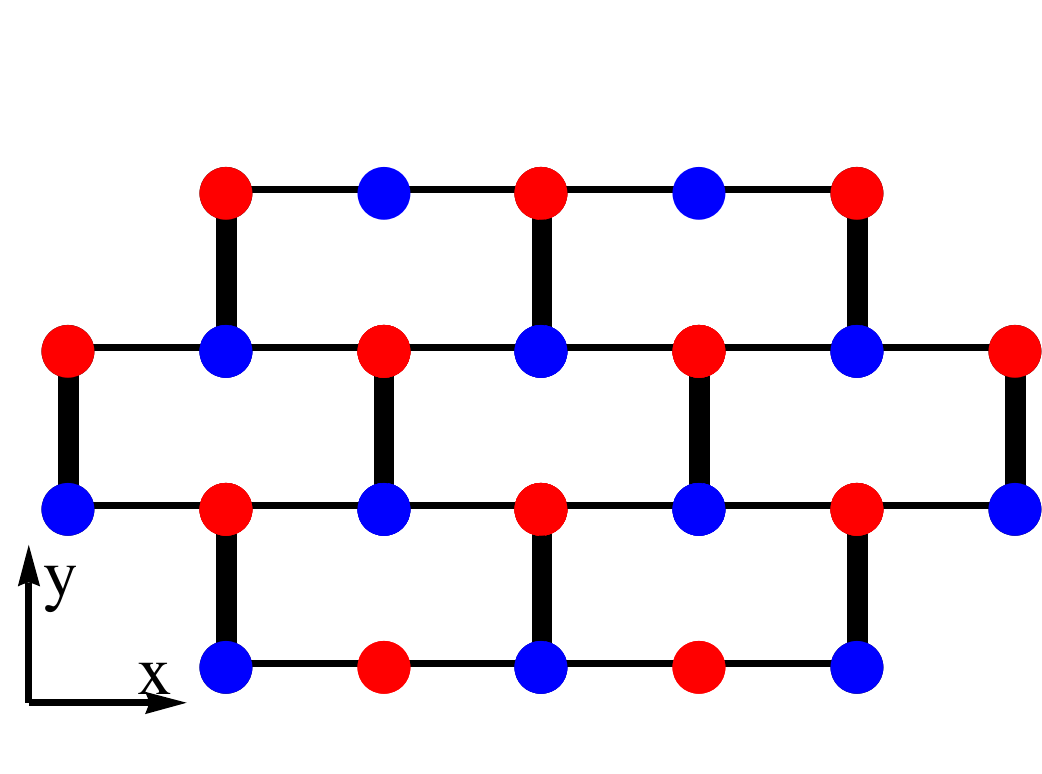}
\includegraphics[width=3.5 cm]{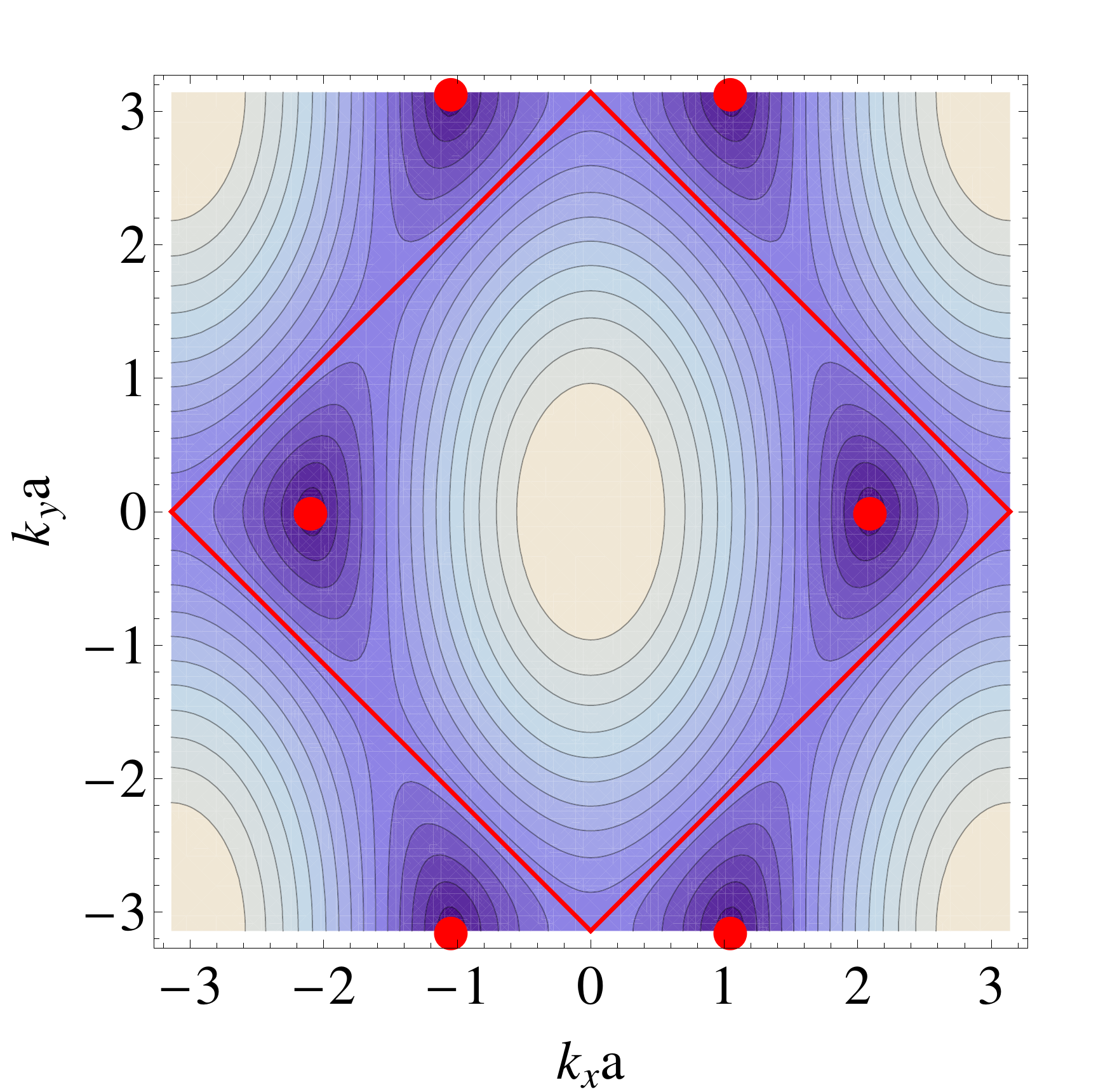}\includegraphics[width=3.5 cm]{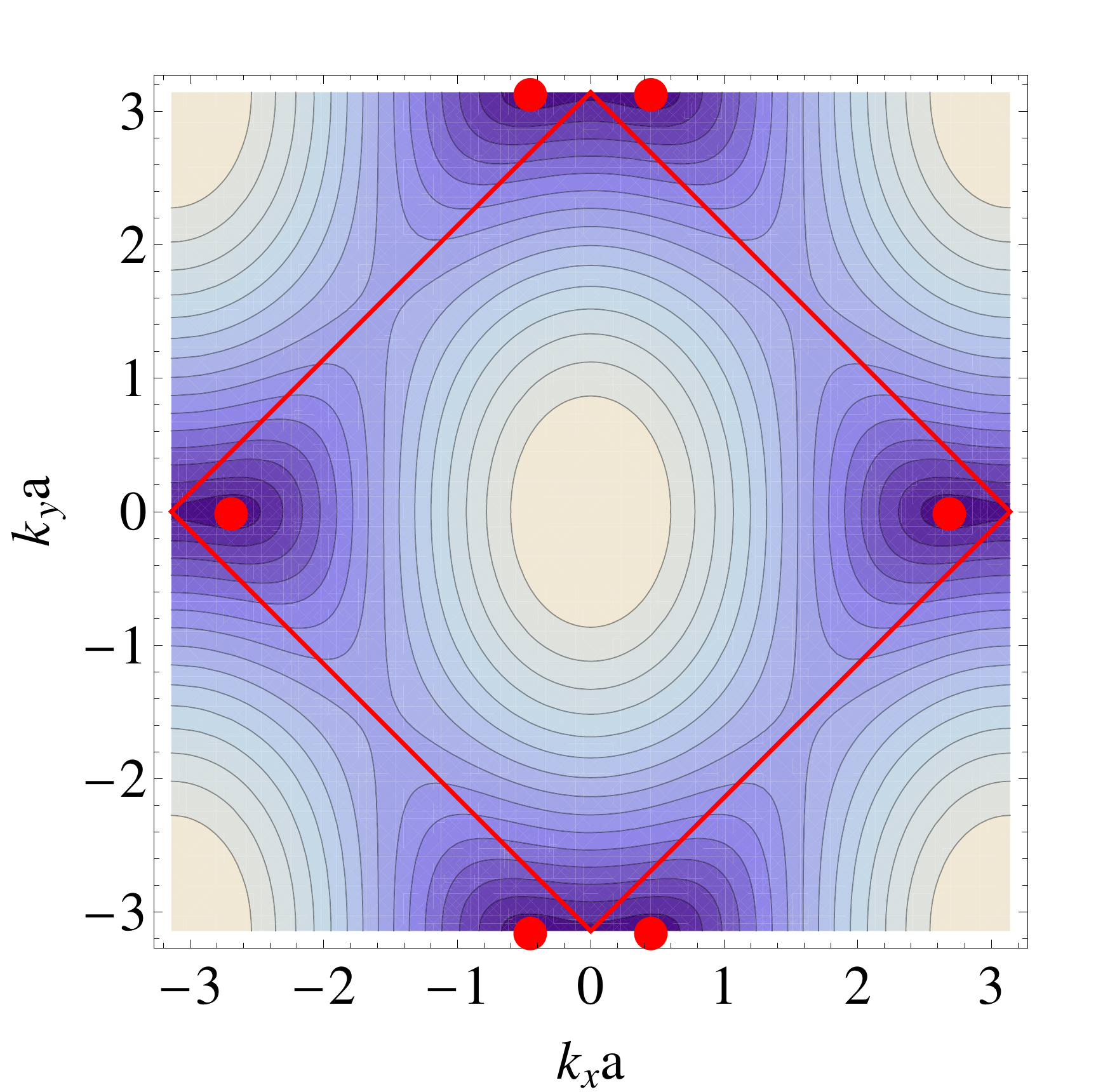}\includegraphics[width=3.5cm]{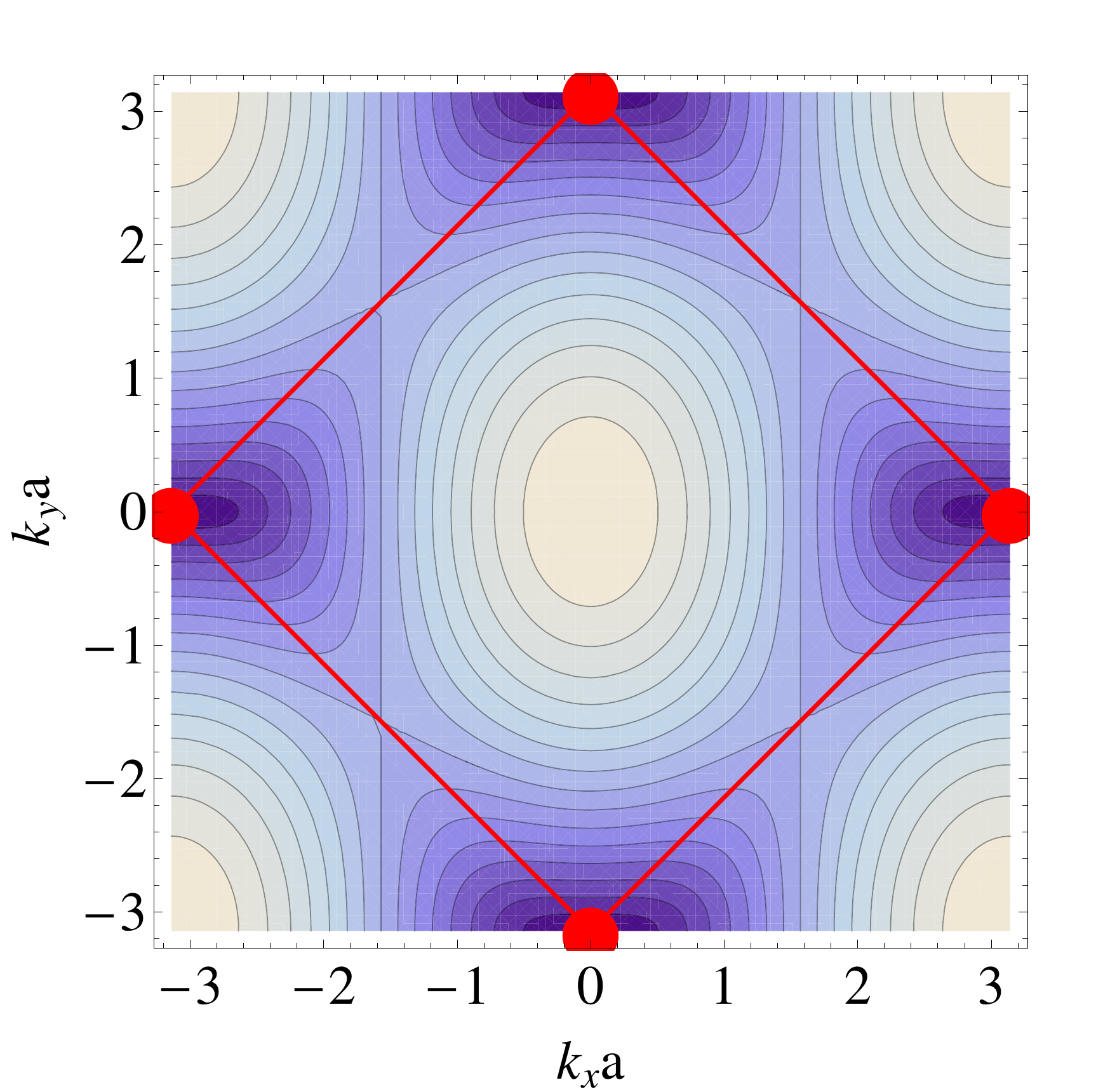}
\end{center}
\caption{Left~: brickwall crystal. The horizontal thick lines indicate the modified hopping integral $t'$. Right~: variation of the isoenergy lines for three values of $t'/t = 1, 1.8, 2$.}
\label{fig:brickwall}
\end{figure}

In the previous sections, we have seen that Dirac points appear as topological objects characterized
by a charge. This charge describes the winding of the wave function in the reciprocal space when
turning around the Dirac point. In this section, we describe how a pair of such Dirac points with
opposite charge related by time-reversal symmetry can move in reciprocal space or even
be annihilated. As mentioned above for the case of the honeycomb lattice relevant in the description
of a graphene crystal, the two Dirac points are precisely located at two opposite corners of the Brillouin zone (Fig.~\ref{fig:honeycomb}).
The position of the Dirac points $\D = \pm (\a_1^* - \a_2^*)/3$ at these high-symmetry points is, however, a rather exceptional situation. Consider for example the ``brickwall'' lattice depicted on Fig.~\ref{fig:brickwall}-a. It has the same couplings between sites as in graphene, but due to the square symmetry, the Brillouin zone is a square. The Dirac points are now located {\it inside} the first Brillouin zone (BZ)  (Fig.~\ref{fig:brickwall}-b). Although this brickwall crystal may appear impossible to realize in condensed matter, it has  recently been realized for a crystal of cold atoms in an optical lattice \cite{Tarruell:12}, as we will discuss later in this review.

 Moreover, upon variation of the band parameters, the two Dirac points may
approach each other and merge into a single point $\DD_0$. This
happens when $\D=-\D$   modulo a reciprocal lattice vector $\G$.
 Therefore, the location of this merging point is
simply $\D_0= \G/2$. There are then {\em  four} possible inequivalent
points whose  coordinates  are $\D_0= ( p\a^*_1 + q \a_2^*)/2$,
with $(p,q)$ = $(0,0)$, $(1,0)$, $(0,1)$,  $(1,1)$.
These are precisely the time-reversal-invariant momenta (TRIM) of a 2D BZ.
 The condition for the existence of Dirac points,
$ f_{\D_0}= \sum_{mn} (-1)^{\beta_{mn}} t_{mn} =0$, where
$\beta_{mn}= p m + q n$, defines a manifold in the space of band
parameters. This manifold separates a
semi-metallic phase with two Dirac cones and a band insulator.
One notices that the merging of Dirac points may even occur at the $\Gamma$ point, under the condition that the hopping parameters do not have the same sign. This sign change may
e.g. be achieved in shaken optical lattices \cite{Koghee}.

Remarkably, at the merging   point, the velocity $\v_1$ vanishes, since  $\sin (\G \cdot \R_{mn}/2)=0$, so that the dispersion becomes massive along this direction, that we define as the  $x$ direction.
This is a direct consequence of the form of the low-energy Hamiltonian in the vicinity of a TRIM.
In order to respect time-reversal symmetry, it must satisfy, in terms of the continuum wave vector
$\q=\D_0-\k$, the same structure
(\ref{eq:TR}) [and (\ref{eq:IR}) in the case of an inversion-symmetric system] as the full
band Hamiltonian (\ref{GH}) around the central $\Gamma$ point.
 Therefore, to lowest order, the   Hamiltonian
 may be expanded as
  \be {\cal H}_0(\q)={q_x^2  \over 2 m^*}\, \sigma^x + c q_y \, \sigma^y \ee
where the velocity $c_y$ and the effective mass $m^*$  may be related to the microsopic parameters (\ref{fofk}) of the original Hamiltonian \cite{Montambaux:09}.
    The terms
of order $q_y^2$ and $q_x q_y$ are neglected at low energy
\cite{Montambaux:09,Dietl}. Most saliently, the corresponding energy spectrum

\be \ep = \pm \left[ c_y^2 q_y^2+\left({q_x^2 \over  2 m^*}\right)^2\right]^{1/2}
\ee
is linear in one direction and quadratic in the other, and the hybrid band-contact point
has also been called a {\it semi-Dirac} point \cite{Banerjee}.

\begin{figure}[htbp!]
\begin{center}
\includegraphics[width=6cm]{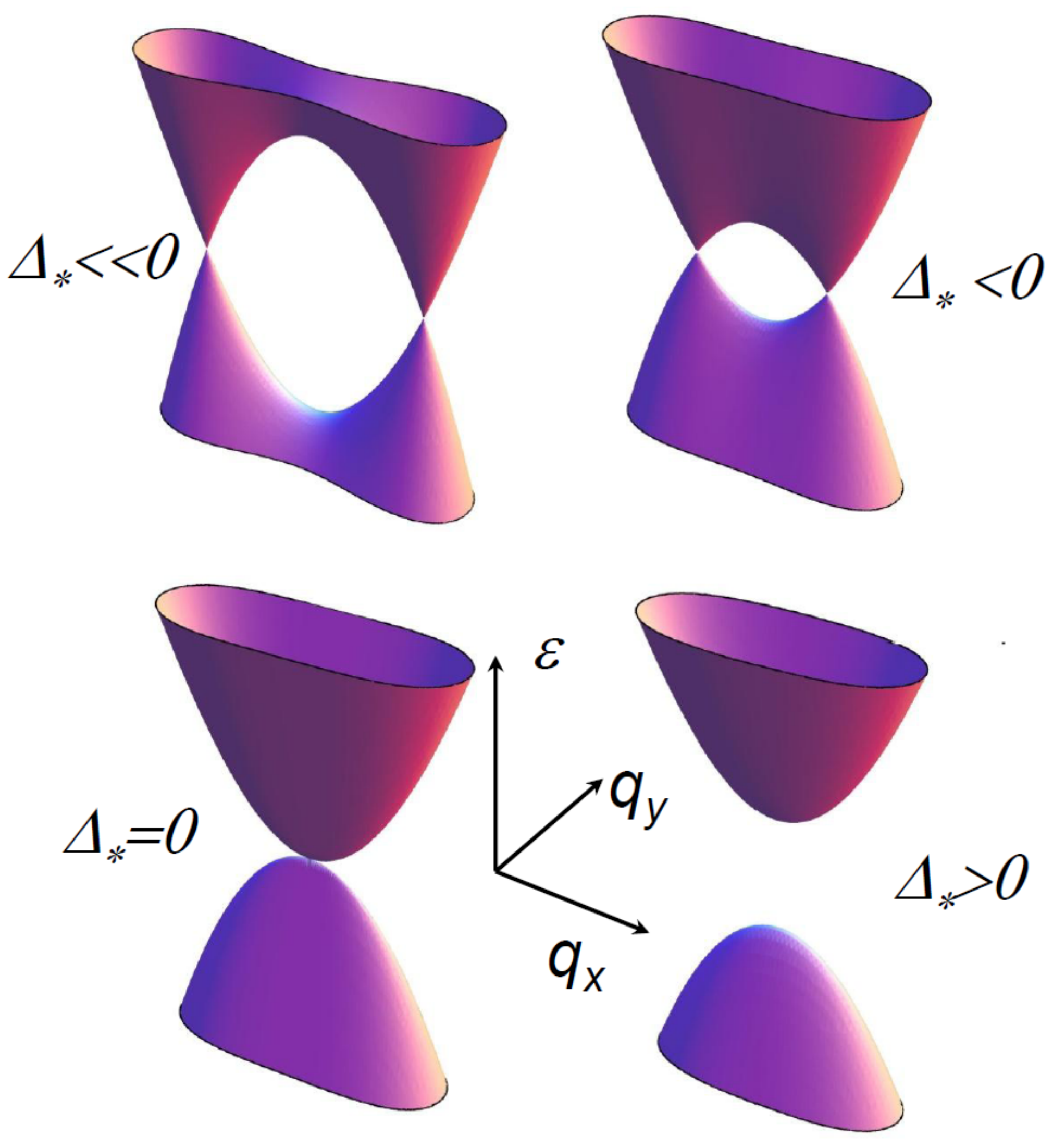}
\includegraphics[width=6cm]{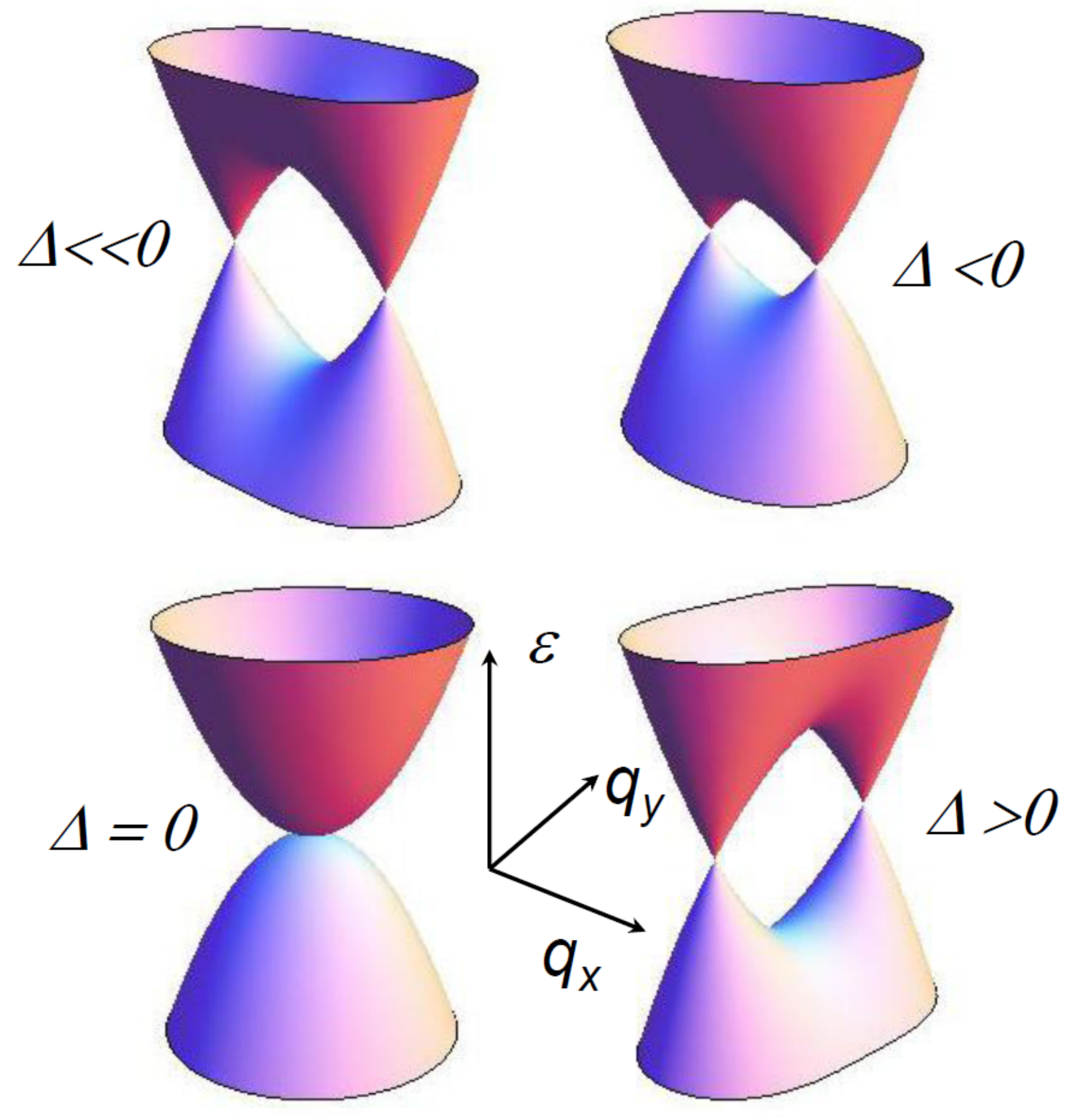}
\end{center}
\caption{Universal scenario for the merging of two Dirac points a) with opposite winding numbers; b) with the same winding number. A gap may open in the first case, but not in the second case. The first case describes the merging of two Dirac points in strained honeycomb lattice. The second case describes the evolution of the spectrum in twisted bilayer graphene (neglecting the trigonal warping (see section \ref{sect:univ2}).}
\label{fig:bicones}
\end{figure}

The merging of the Dirac points in $\DD_0$ marks the transition
between a semi-metallic phase and an insulating phase, and it can be analysed topologically in
terms of winding numbers -- below the transition, the semi-metallic phase is characterised by
two Dirac points of opposite charge that are annihilated at the transition. The resulting
zero topological charge then allows for the opening of a gap in the spectrum and a transition
to an insulating phase. In order to describe the transiton more quantitatively, we introduce
the gap parameter

\be \Delta_*= f_{\D_0}= \sum_{mn} (-1)^{\beta_{mn}} t_{mn} \label{gap} \ee
which changes  its sign at the transition.  In the vicinity of the transition,
the Hamiltonian has the universal form

\be {\cal H}_{+-}(\q)= 
\left(
  \begin{array}{cc}
    0 & \Delta_*+ {q_x^2 \over 2 m^*} -  i c_y q_y  \\
 \Delta_* + {q_x^2 \over 2 m^*} +  i c_y q_y & 0 \\
  \end{array}
\right) \label{newH} \ee
with the spectrum
\be
\ep= \pm \sqrt{\left(\Delta_* + {q_x^2 \over 2
m^*}\right)^2 + q_y^2 c_y^2} \ . \label{Ueps1}
\ee
This \textit{Universal Hamiltonian}, which describes the merging of two Dirac points
with opposite charge \cite{Montambaux:09}, is universal in the sense that its structure is
general, independent of the microscopic parameters.
It has a remarkable structure
and describes properly the vicinity of the topological transition,
as shown on Fig. \ref{fig:bicones}-a. When $ \Delta_*$ is negative
(we  choose $m^* >0$ without loss of generality), the spectrum
exhibits the two Dirac cones at a distance $2 q_D = 2\sqrt{- 2 m^* \Delta_*}$ and a saddle point in $\D_0$ (the energy of the saddle point being $\pm |\Delta_*|$). Increasing $\Delta_*$ from
negative to positive values, the saddle point shifts to lower energies and eventually disappears
into the hybrid semi-Dirac point at the transition   ($\Delta_*=0$), before
 a gap of size $2 \Delta_* >0$ occurs in the spectrum.

 Therefore, the Hamiltonian can also be seen as an interpolation between the behaviour of isolated Dirac points like in graphene to massive particles in the gapped phase. In particular, the Landau level spectrum in a magnetic field evolves continuously from the well-known  $\sqrt{n B}$ spectrum as in graphene to a massive particules spectrum $(n+1/2) B$ above the merging transition, as one may
see from the semi-classical quantisation rule (\ref{SCQR}). Whereas below the transition there exist,
at sufficiently low magnetic fields, closed orbits encircling just one of the Dirac points (with
a winding number $|w_i|=1$) and one therefore obtains a doubly degenerate zero-energy level, the
situation is drastically different above the transition. Indeed, all possible orbits then have
a winding number $w=0$ that yields the $1/2$ offset in the Landau-level spectrum, as in the case
of conventional Schr\"odinger fermions.
Directly at the transition, the level spectrum shows an unusual behaviour $[(n+1/2) B]^{2/3}$.
Again the $1/2$ offset is due to the absence of closed orbits encircling singular points with
$w\neq 0$ at the transition. Notice, however, that the topological transition in the presence
of a magnetic field is not abrupt. Also below the transition, where the zero-field spectrum reveals
two Dirac points, the two Dirac points are coupled by the magnetic field -- indeed, the closed orbits
necessarily enclose surfaces of size $1/l_B^2\propto B$
in reciprocal space, due to the non-commutativity
of the kinetic-moment operators (\ref{eq:Peierls}), and, at sufficiently large magnetic fields,
Dirac points separated by small wave vectors are no longer resolved. This lifts the original
two-fold degeneracy of the zero-energy level $n=0$ in an exponential manner.
The continuous evolution of this level is discussed in Ref. \cite{Dietl}.

As a simple example, the motion and merging Dirac points may be realized in the above honeycomb and brickwall lattices with first nearest neighbors coupling, where one of the coupling parameters named $t'$ has been increased \cite{remark1}

\be f_{\k} =   t (\beta + e^{i \k \cdot \a_1} + e^{i \k \cdot \a_2} )  \qquad, \qquad \mbox{with} \ \beta= t'/t \ee
with $\a_i= (\pm {\sqrt{3} \over 2} a , {3 \over 2} a)$ for the honeycomb lattice and $\a_i=(\pm a, a)$ for the brickwall lattice, $a$ being the interatomic distance.
The parameters of the Universal Hamiltonian are then respectively $ \Delta_* = t'-2 t, m^*=2/(3t), c_y = 3 t$
and
$ \Delta_* = t'-2 t, m^*=1/(2t), c_y =   t$. The merging scenario initially proposed  in elongated graphene \cite{Montambaux:09,Dietl,Hasegawa1,Guinea:08} turned out to be unreachable \cite{Pereira},
but it has been observed in different systems, now called ``artificial graphenes'', that
we discuss in the following section.

\section{Manipulation of Dirac points in artificial graphenes}

The intensive study of Dirac fermions in
graphene has motivated the search for different systems sharing similar properties with graphene,  in particular to exhibit phenomena which could not be observed in graphene. The flexibility of such systems may allow for  the realization of properties unreachable in graphene, like the predicted topological transition or the manipulation of edge states. Examples for these artificial graphene comprise
photonic or microwave crystals \cite{Bellec:13a,Segev:13,Kuhl,Richter},
molecular crystal \cite{Manoharan:12}, ultracold atoms in optical lattices \cite{Tarruell:12}, polaritons propagating in a honeycomb lattice of  coupled micropillars etched
in a planar semiconductor microcavity \cite{Amo}, or  the quasi-2D organic salt $\alpha$-(BEDT-TTF)$_2$I$_3$ under pressure \cite{ET3}. We do not elaborate further on this now long list of different physical systems (for a review, see Ref. \cite{Polini:13}), but here we restrict the discussion to
only two physical systems where the manipulation and merging of Dirac points has been explicitly observed and studied.

\subsection{A lattice of cold atoms}

\begin{figure}[htbp!]
\begin{center}
\includegraphics[height=8cm]{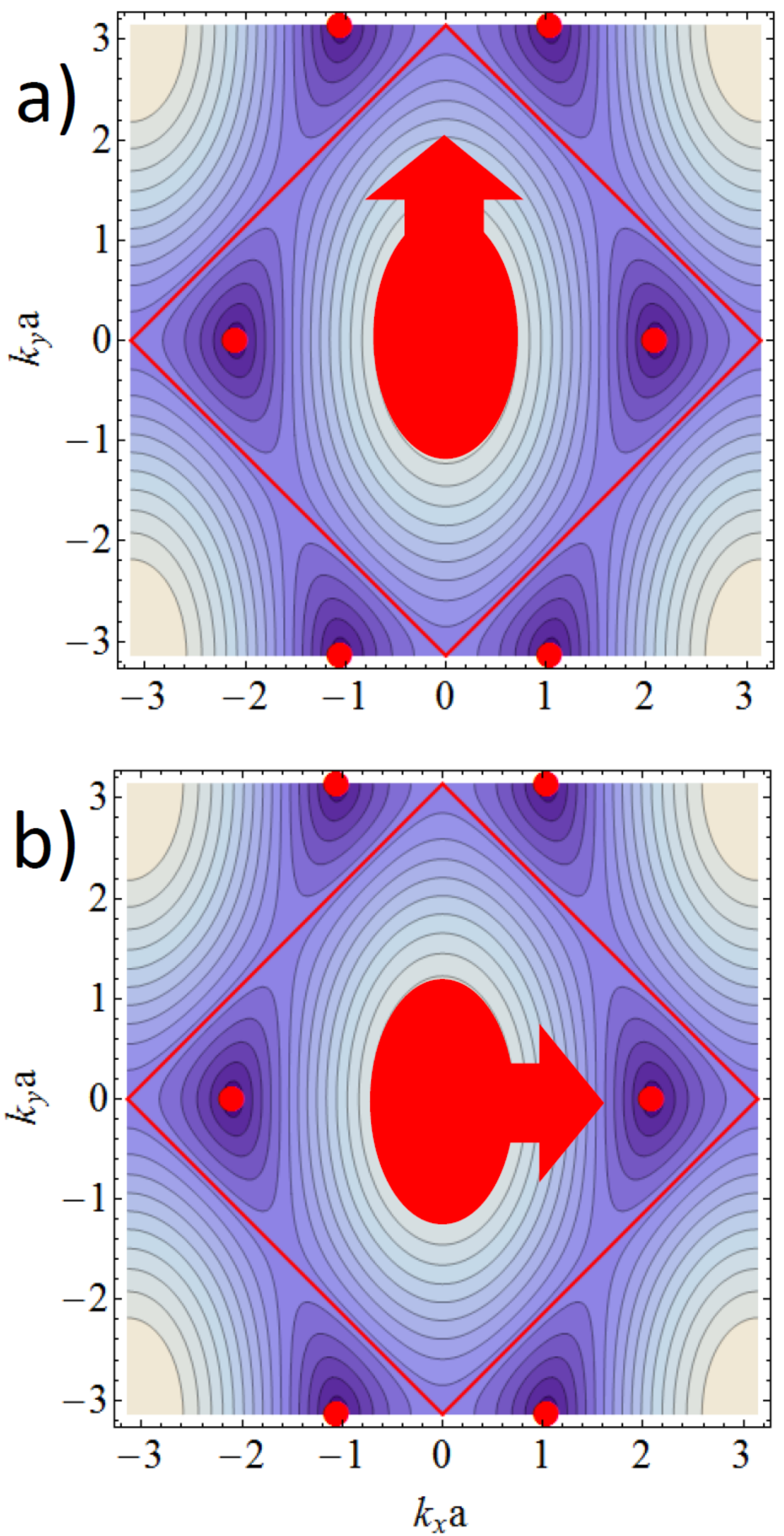}
\includegraphics[height=8cm]{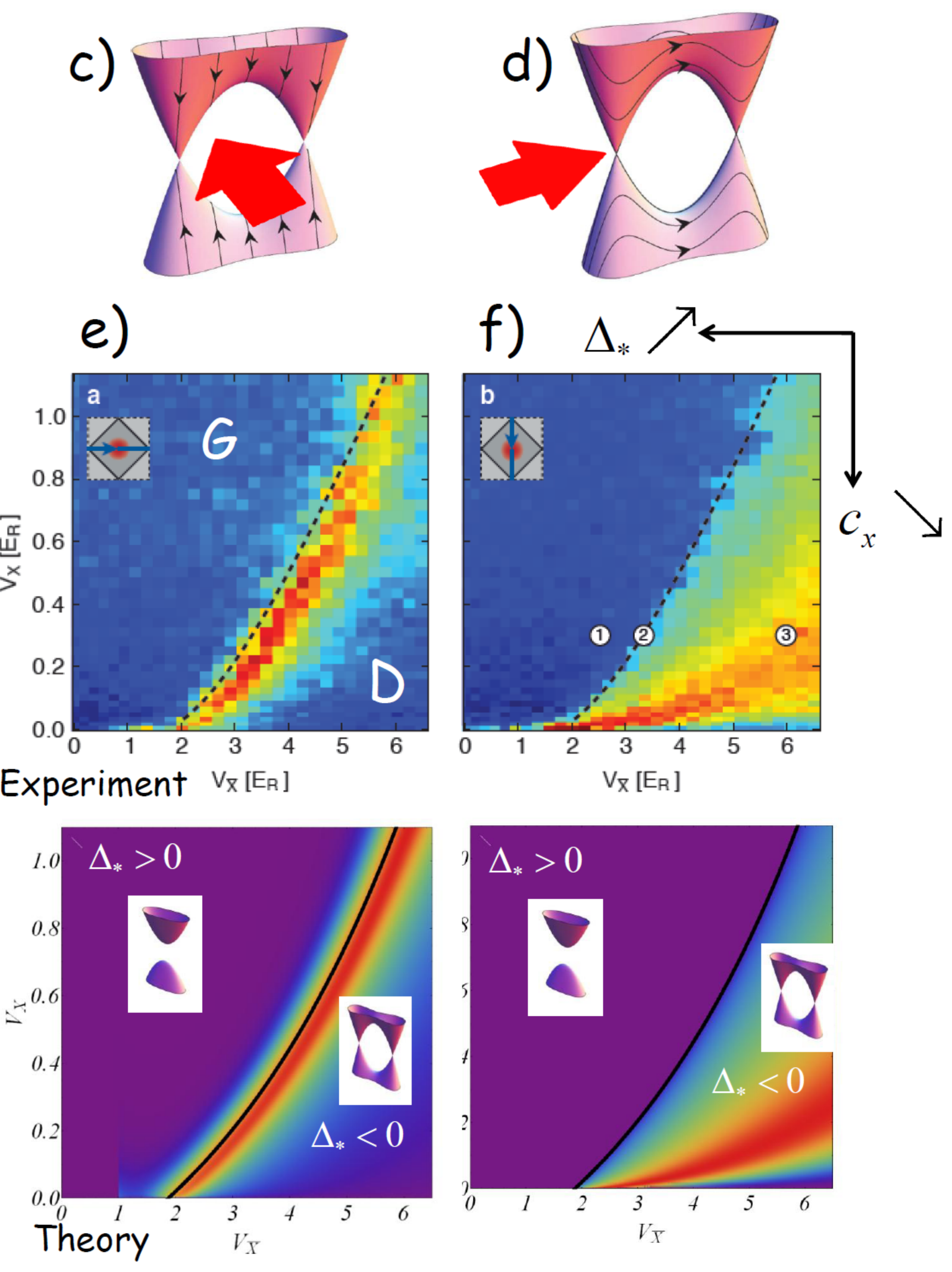}
\end{center}
\caption{a,b) Schematic picture of the low energy fermionic cloud in reciprocal space. In case a,c), the force is applied in a direction perpendicular to the merging direction so that the fermionic cloud ``hits'' the two Dirac points in parallel. In case b,d), the force is applied along the merging direction to that the could encounters the tow Dirac points in series, leading to a double Landau-Zener tunneling.}
\label{fig:LZ-exp-theo}
\end{figure}

Ultracold atoms trapped in an optical lattice offer beautiful realizations of condensed matter situations. It has recently been possible to create a periodic potential with the help of
standing optical (laser) waves that trap cooled atoms via a dipolar interaction. These trapped atoms
can be described to great accuracy
within a tight-binding model simulating very closely the physics of graphene. The lattice is indeed very close to the brickwall lattice depicted on Fig. \ref{fig:brickwall}. By varying appropriately the intensities of the laser fields, it is possible to realize exactly the merging scenario described by the Universal Hamiltonian (\ref{newH})
\cite{Montambaux:09,Lim:12}. In order to probe the spectrum, the position of the Dirac points, their motion and their merging, a low-energy cloud of fermionic atoms is submitted to a constant force $F$ (Fig.  \ref{fig:LZ-exp-theo}-a,b), so that its motion is uniform in reciprocal space $\hbar (d \k/ d t) = \F$, and exhibits Bloch oscillations \cite{Tarruell:12}.  In the vicinity of a Dirac point, there is a finite probability for the atoms to tunnel into the upper band. This probability depends on the applied force and on the gap separating the two bands. For a single crossing, it is given by
Landau-Zener (LZ) theory \cite{Zener}. By measuring the proportion of atoms having tunneled into the upper band after one Bloch oscillation, it is in principle possible to reveal the energy spectrum. Since the spectrum exhibits a {\it pair} of Dirac points, it is important to separate two cases, as it has been done experimentally.\footnote{Here we define direction $x$ and $y$ consistent with the rest of the paper. They are interchanged compared to Refs. \cite{Tarruell:12,Lim:12}}
\medskip

{\it -- Single LZ tunneling.}  In this case, the force is applied along the $y$-direction perpendicular to the merging line and  the cloud of atoms ``hits'' the two Dirac points in parallel  (Fig.  \ref{fig:LZ-exp-theo}-a,c).  An atom, initially in a state with finite $q_x$, performs a Bloch oscillation along a line of constant $q_x$ and may tunnel into the upper band with a probability  given by
\be P_Z^y=  e^{ \displaystyle -\pi  {(\textrm{gap/2})^2 \over  c_y F}}=e^{ \displaystyle -\pi  {({q_x^2 \over 2 m^*} + \Delta_* )^2 \over  c_y F}} \label{pzy} \ee
where we have written the gap in terms of the  parameters of the Universal Hamiltonian (\ref{newH}). In the gapped (G) phase, above the merging transition ($\Delta_* >0$), the tunneling probability is vanishingly small. In the opposite case, deep in the Dirac (D) phase ($\Delta_* <0$), when the distance $2 q_D= 2 \sqrt{- 2 m^* \Delta_*}$ between the Dirac points is larger than the size of the cloud, the probability to  tunnel into the upper band is also small.  The tunnel probability is actually large near the merging transition.  Figure \ref{fig:LZ-exp-theo}-e represents the tunneling probability as a function of two parameters of the lattice potential ($V_X,V_{\overline X}$) that we do not explicit here \cite{Tarruell:12}.  This result has been obtained by relating these parameters to the parameters of the Universal Hamiltonian  and by using Eq. (\ref{pzy}) \cite{Lim:12}. The agreement with the experiment is excellent without any adjustable parameter (Figs. \ref{fig:LZ-exp-theo}-e). To account quantitatively for the experimental result, one has to average the probability (\ref{pzx}) over a finite range of $q_x$, due to the finite size of the fermionic cloud in reciprocal space   (Fig. \ref{fig:LZ-exp-theo}-a). By doing so, one sees from (\ref{pzy}) that the probability is maximal for a negative value of the driving parameter $\Delta_* = - \langle q_x^2 \rangle /2 m^* $ where $\langle \cdots \rangle$ is an appropriate average. This explains why the intensity is maximal {\it inside} the D phase, as seen in Figs. \ref{fig:LZ-exp-theo}-e.

\medskip

{\it -- Double LZ tunneling.}  In this case the force $F$ is applied along the merging $x$-direction  and the cloud of atoms ``hits'' the two Dirac points in series (Fig.  \ref{fig:LZ-exp-theo}-b,d). This situation is more involved since each atom may undergo two LZ transitions in a row. Each LZ transition is described by the tunnel probability
\be P_Z^x = e^{ \displaystyle -\pi  {(\textrm{gap/2})^2 \over  c_x F}}= e^{\displaystyle -   \pi {c_y^2 q_y^2  \over c_x F}}= e^{\displaystyle -   \pi {c_y^2 q_y^2  \over  F \sqrt{2|\Delta_*|/m^*}}} \ .  \label{pzx} \ee
Assuming that the two tunneling events are incoherent,  the interband transition probability resulting from the two event in series is
\be P_t^x = 2 P_Z^x (1 - P_Z^x) \ .  \label{Ptx} \ee

In the G phase, the tunneling probability is again vanishingly small.
In the D phase, when    $q_y=0$, the single LZ probability is maximal ($P_Z^x=1$), but the tunneling probability after two events vanishes. For an initial cloud of finite size $q_y$, the transferred fraction is an average $\langle P_t^x \rangle$ taken on the finite width of the cloud.
 The interband transition probability [Eq. (\ref{Ptx})] is a non-monotonic function of the LZ probability $P_Z^x$, and it is  maximal when $P_Z^x=1/2$.  This explains why the maximum of the tunnel probability  is located well {\it inside} the D phase (red region in Figs.  \ref{fig:LZ-exp-theo}-f). Varying the averaging order, this happens when $\langle P_t^x \rangle \simeq 1/2$, that is for a finite value $\Delta_*$ given by     $ m^* c_y F \ln 2 /(2 \pi \langle q_x^2\rangle)$, that is well inside the Dirac phase, as shown in Figs. \ref{fig:LZ-exp-theo}-f.

 A particularly interesting effect related to the double LZ tunneling is the possible existence of interference effects between the two LZ events. Assuming the phase coherence is preserved, instead of
the probability given by Eq. (\ref{Ptx}), one expects a resulting probability of the form

\be P_t^x = 4 P_Z^x (1 - P_Z^x)  \cos^2 (\varphi/2 + \varphi_d) \ ,  \label{Ptxcoh} \ee
where $\varphi_d$ is  a phase delay, named Stokes phase, attached to the each tunneling event, and $\varphi=\varphi_{dyn}+\varphi_{g}$ is a phase which has two contributions,
a dynamical phase $\varphi_{dyn}$ acquired between the two tunneling events  and basically related to the energy difference between the two energy paths, and a geometric phase $\varphi_{g}$. Whereas the dynamical phase carries information about the spectrum, the geometric phase carries information about the structure of the wave functions \cite{Lim:13}. It is now a experimental challenge to access directly this interference pattern and to probe the different contributions to the dephasing.

\subsection{Propagation of microwaves}

\begin{figure}[htbp!]
\begin{center}
\includegraphics[width=10cm]{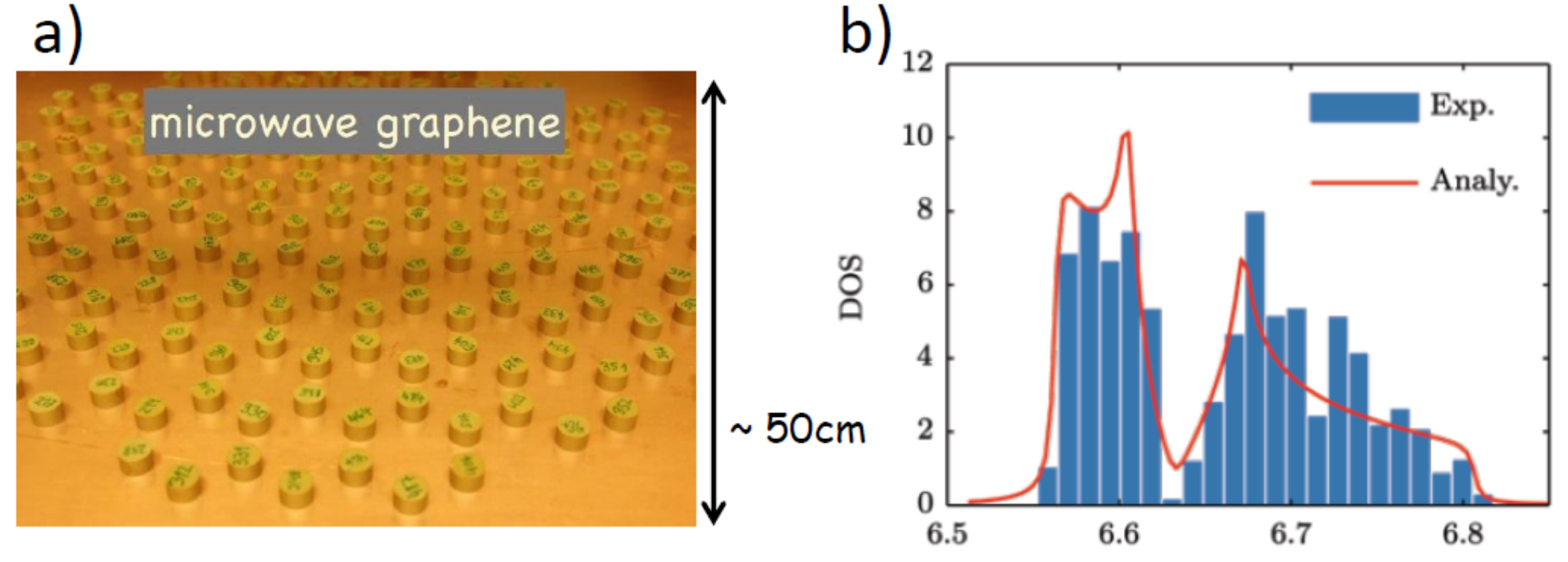}
\includegraphics[width=10cm]{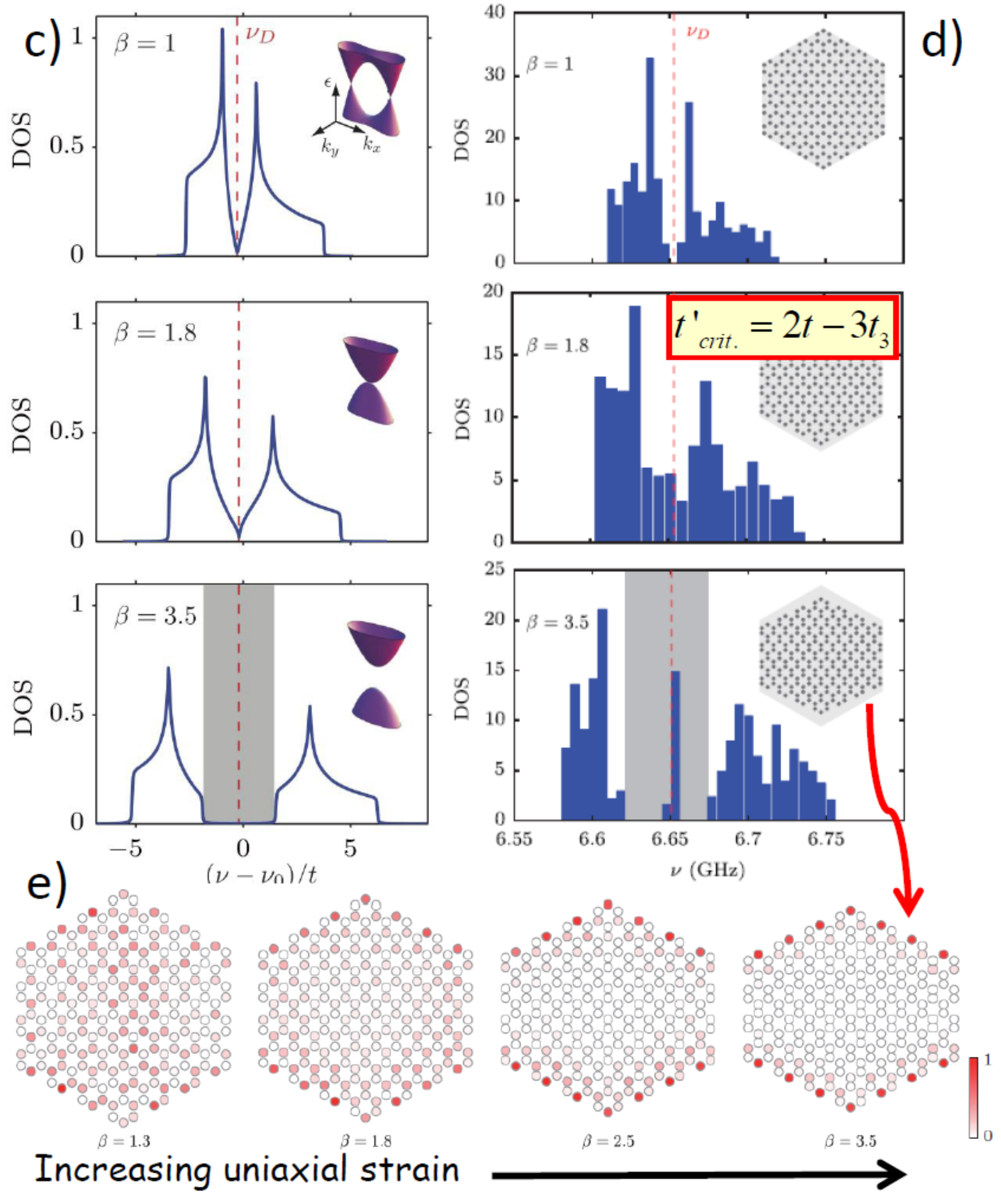}
\end{center}
\caption{a) Honeycomb lattice of 288 dielectric cylinders. b) Experimental DOS well fitted by a tight-binding model with second and third nearest neighbor couplings. c) Expected evolution of the DOS with anisotropy of the hopping parameters $t' \neq t$. d) Experimental evolution of the DOS with a uniaxial deformation of a honeycomb lattice with armchair boundaries. The merging transition occurs for a critical value $t'= 2 t - 3 t_3$. Under strain, new edge states appear at the band centre. e) These new edge states are located at the edges which are not parallel to the strain axis.}
\label{fig:microwaves}
\end{figure}

The rich physical properties associated with the propagation of electrons in a honeycomb lattice may also be revealed in  the propagation of  {\it any} wave in this lattice. Therefore electrons may be  replaced by other waves such as light, microwaves, other elementary excitations like polaritons. This may allow for a much more flexible realization of the same physics, but implying different length scales. As example,  we consider here a microwave that is confined between two metallic plates realizing a 2D situation and that propagates through an ensemble of dielectric cylindric dots of centimeter size (Fig. \ref{fig:microwaves}-a). The frequency is chosen such that the propagation is resonant inside a dot and evanescent outside the dots \cite{Kuhl,Bellec:13a}. Therefore the dots are weakly coupled through evanescent waves and the wave propagation between the dots is very well described within a tight-binding model \cite{Bellec:13a,Bellec:13b}. The signal is emitted and measured by an antenna which gives direct access to the local density of states (DOS).
The measured DOS is plotted in Fig. \ref{fig:microwaves}-b, and it can be described within a tight-binding model where second and third nearest neighbor couplings are not small (they depend on   the distance between the dots and typically  $t_2/t \simeq 0.09, t_3/t \simeq 0.07$). A uniaxial strain is easily realized in this setup, so that one of the coupling $t'$ may be modified and typically the ratio $\beta= t'/t$ has been varied between $0.4$ and $3.5$. By doing so, the merging transition has been reached for a critical value $\beta_{cr}  \simeq 1.8$ which corresponds very well  the theoretically expected value taking into account the higher order nearest neighbor couplings  $\beta_{cr}= 2 - 3 t_3/t $, since from Eq. (\ref {gap}), we have here $\Delta_*=t'-2 t + 3 t_3$.

The great advantage of this setup is its flexibility.   It is quite easy to manipulate the ``atoms'' and to measure the local DOS. This flexibility  has been used to modify at will the structure of the edges and to investigate the existence of edge states whose importance is well-known in graphene.
Indeed, zigzag edges support edge states while armchair edges do not.  It has been predicted however that edge states may exist even in the armchair case, in the presence of uniaxial anisotropy \cite{Delplace:11}. The existence of edge states is clearly revealed by DOS measurement as seen on Fig. \ref{fig:microwaves}-d when the anisotropy increases. It has been found that
(i) edge states appear only along the edges that are
not parallel to the anisotropy axis (Fig. \ref{fig:microwaves}-e). (ii) Their localization
along the edge increases when $\beta$ increases. (iii) Their
existence is not related to the topological transition: they appear as soon as $\beta >1$.
Moreover, it is found that (i) the intensity on one triangular
sublattice stays zero, and (ii) the intensity on the other
sublattice decreases roughly as $1/\beta^{2r}$, where $r$ is the
distance to the edge in units of the lattice parameter. These
features are in agreement with the prediction for the existence
of armchair edge states in deformed structures, and the existence of edge states has been related to a topological property of the bulk wave functions, the Zak phase \cite{Delplace:11}.
More extensive investigation of these states, as well as of
the states along zig-zag and bearded edges in anisotropic structures, is in progress \cite{Bellec:14}.

\section{More Dirac points}
\label{sec:NDP}

In the framework of the general tight-binding model (\ref{GH}), one can also be confronted
with situations where there are several pairs of Dirac points. The generation and motion of these
additional Dirac points, as well as the possible fusion of Dirac points with like topological charge,
are the issue of the present section.

\subsection{Monolayer with third neighbor coupling}
\label{sect:tt3}

\begin{figure}[h!]
\begin{center}
\includegraphics[width=4.5cm]{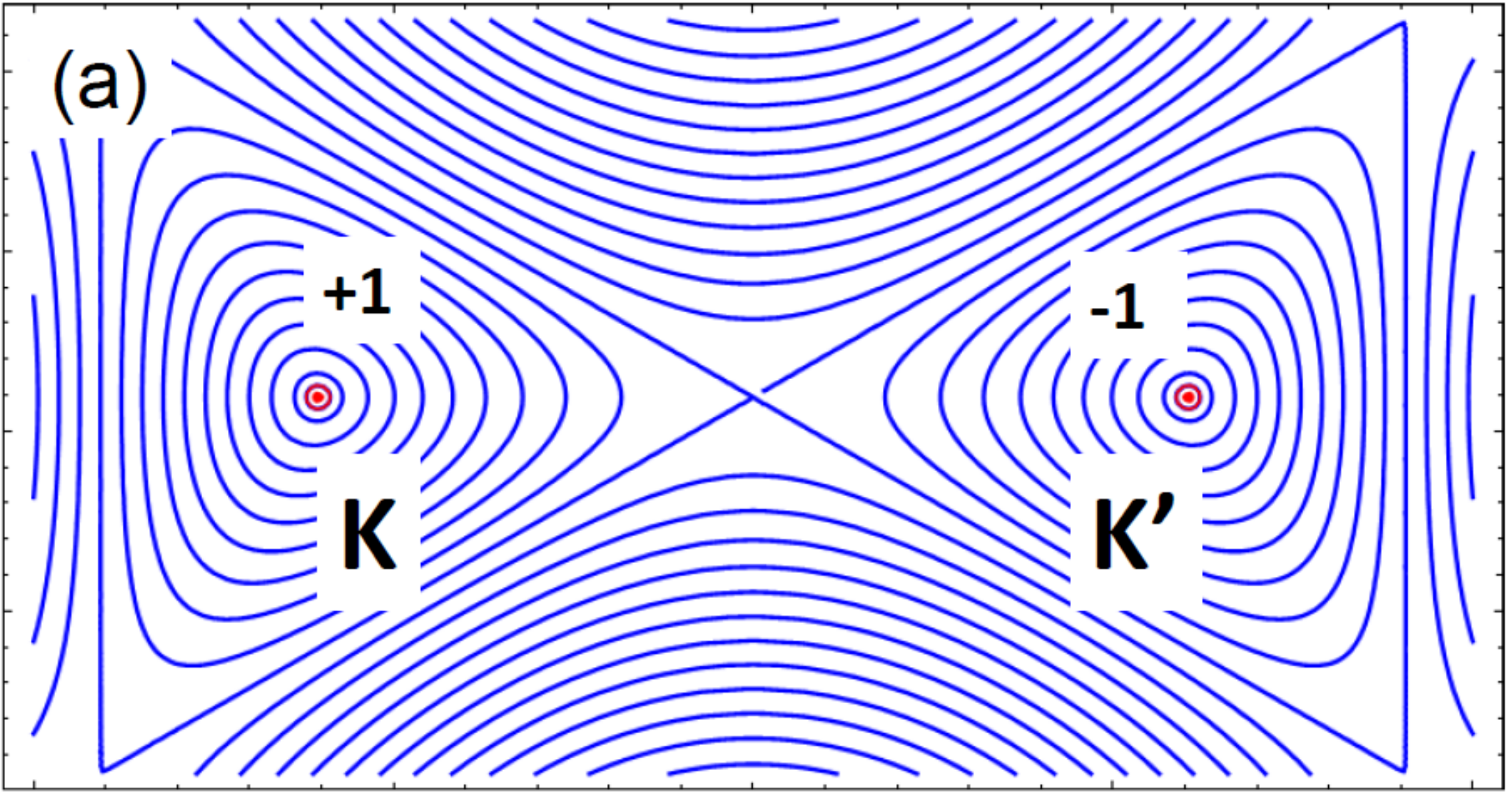}\includegraphics[width=4.5cm]{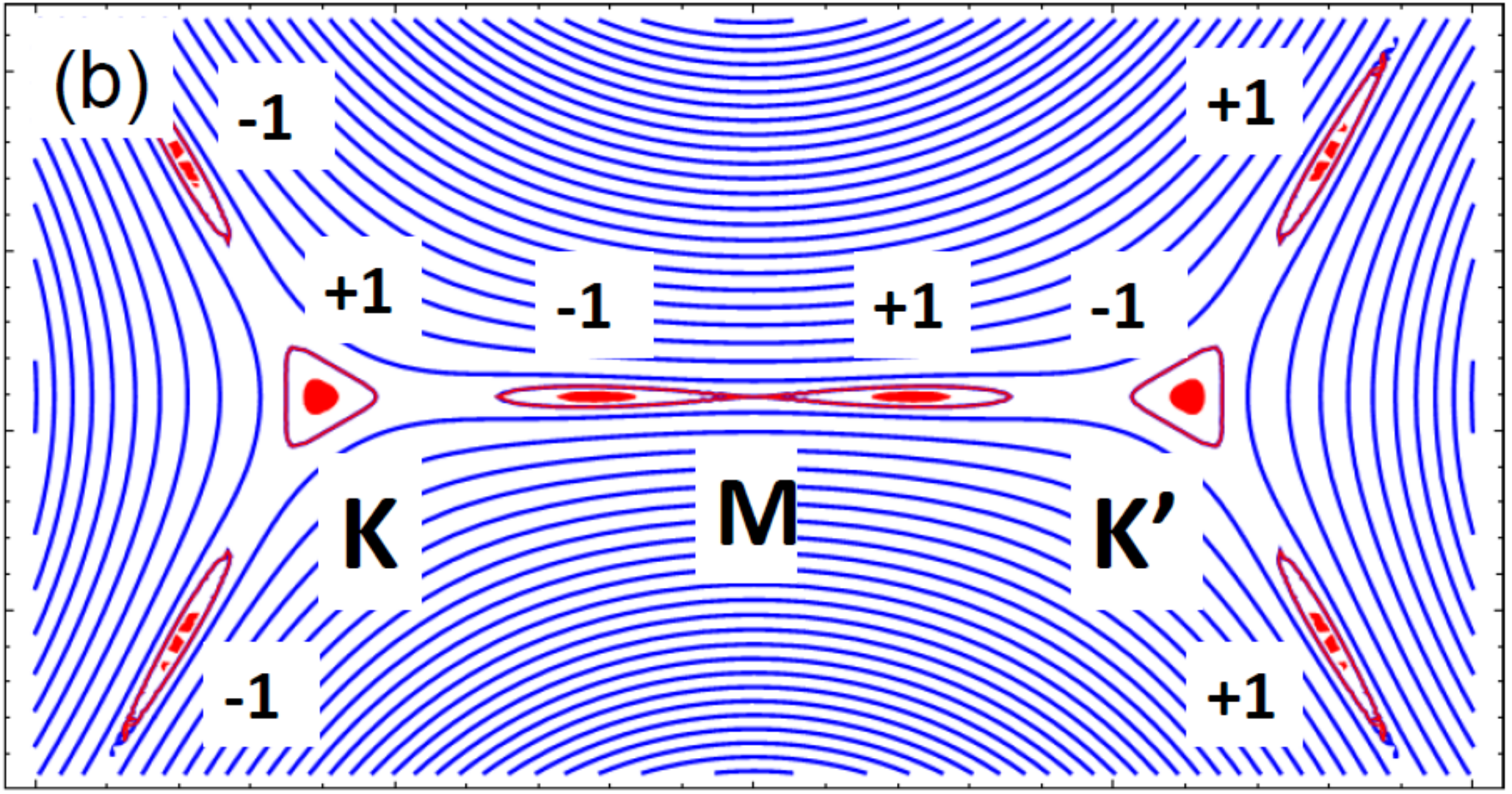}
\end{center}
\begin{center}
\includegraphics[width=4.5cm]{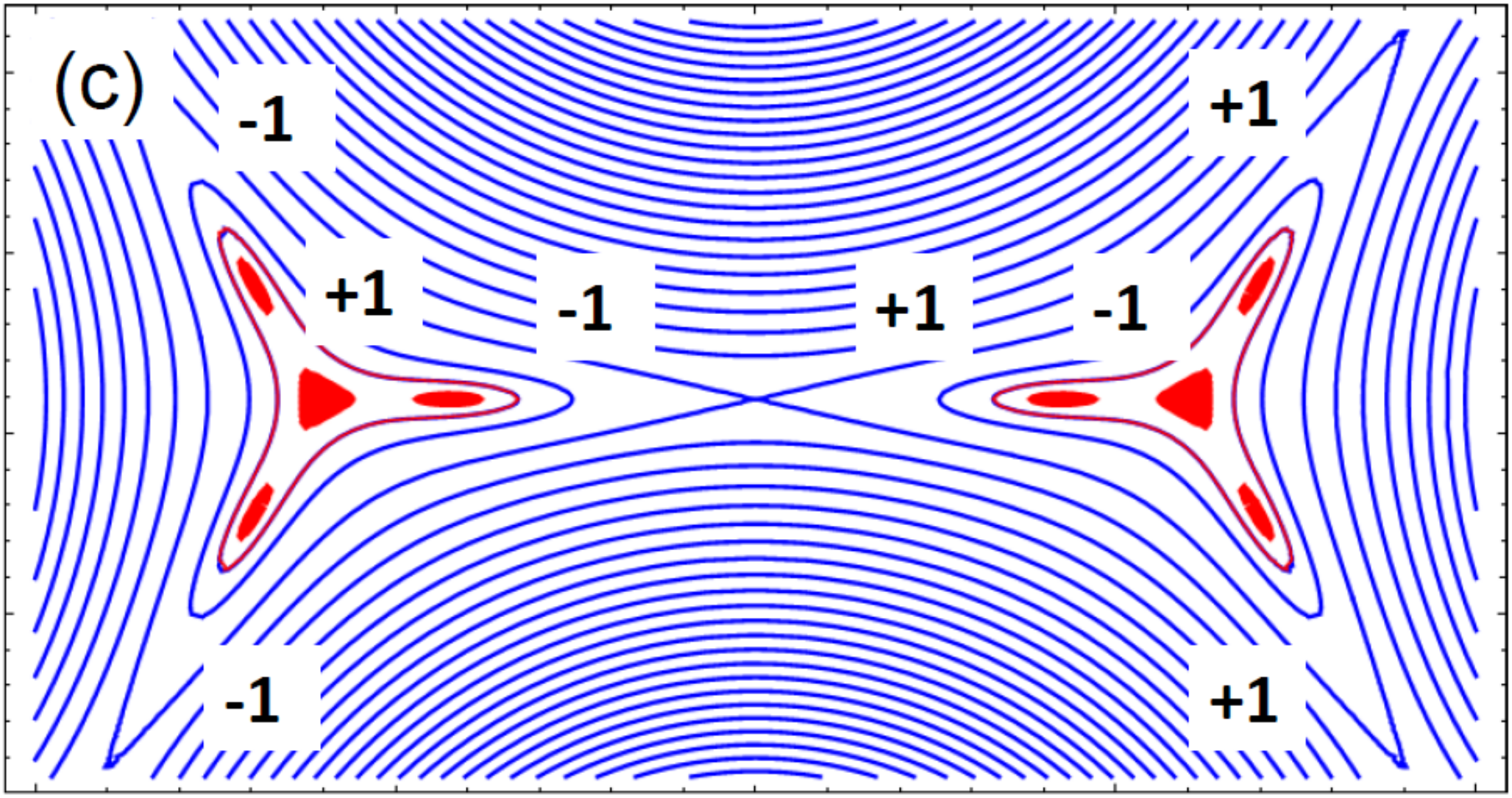}\includegraphics[width=4.5cm]{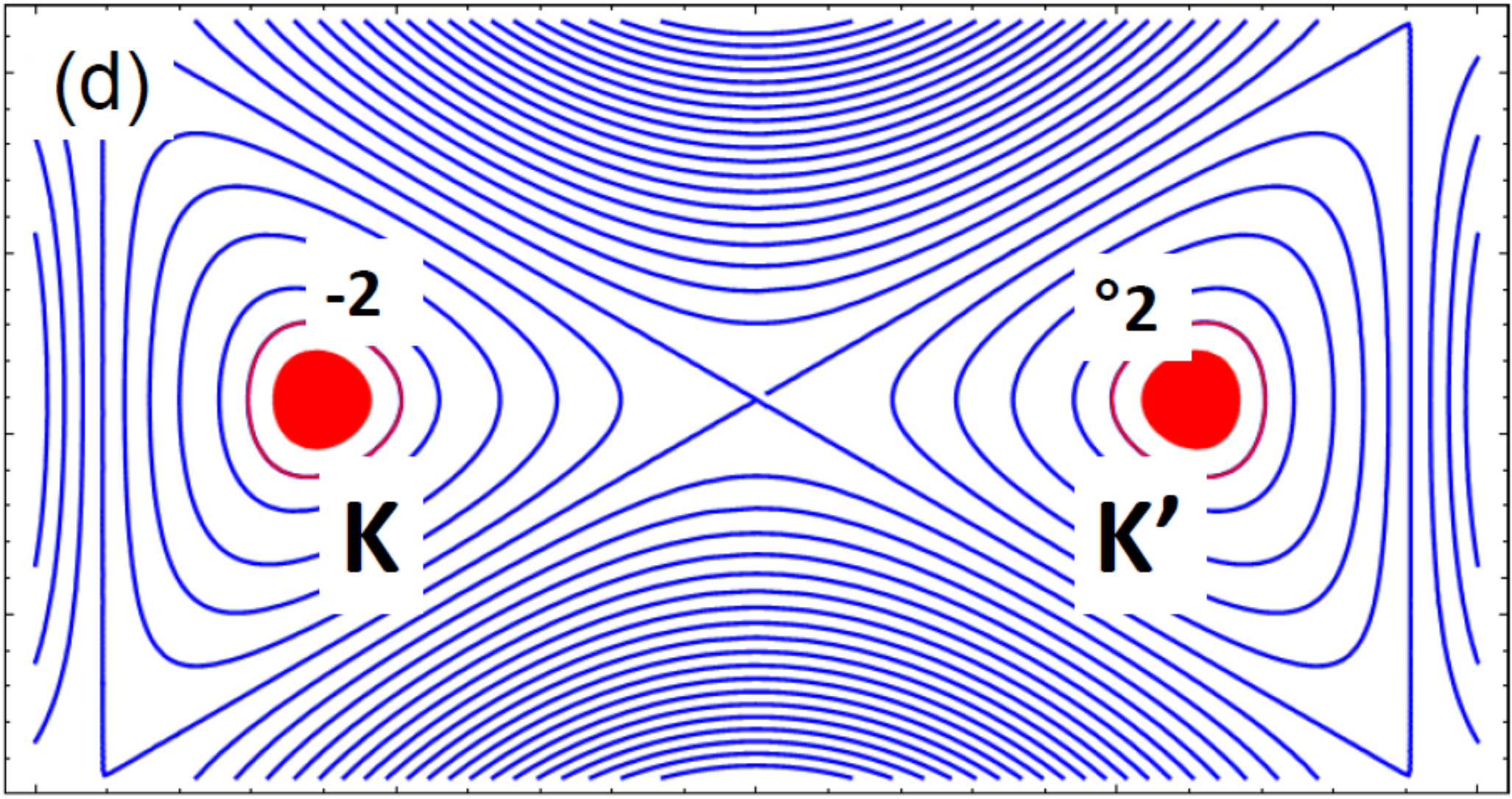}
\end{center}
\begin{center}
\includegraphics[width=4.5cm]{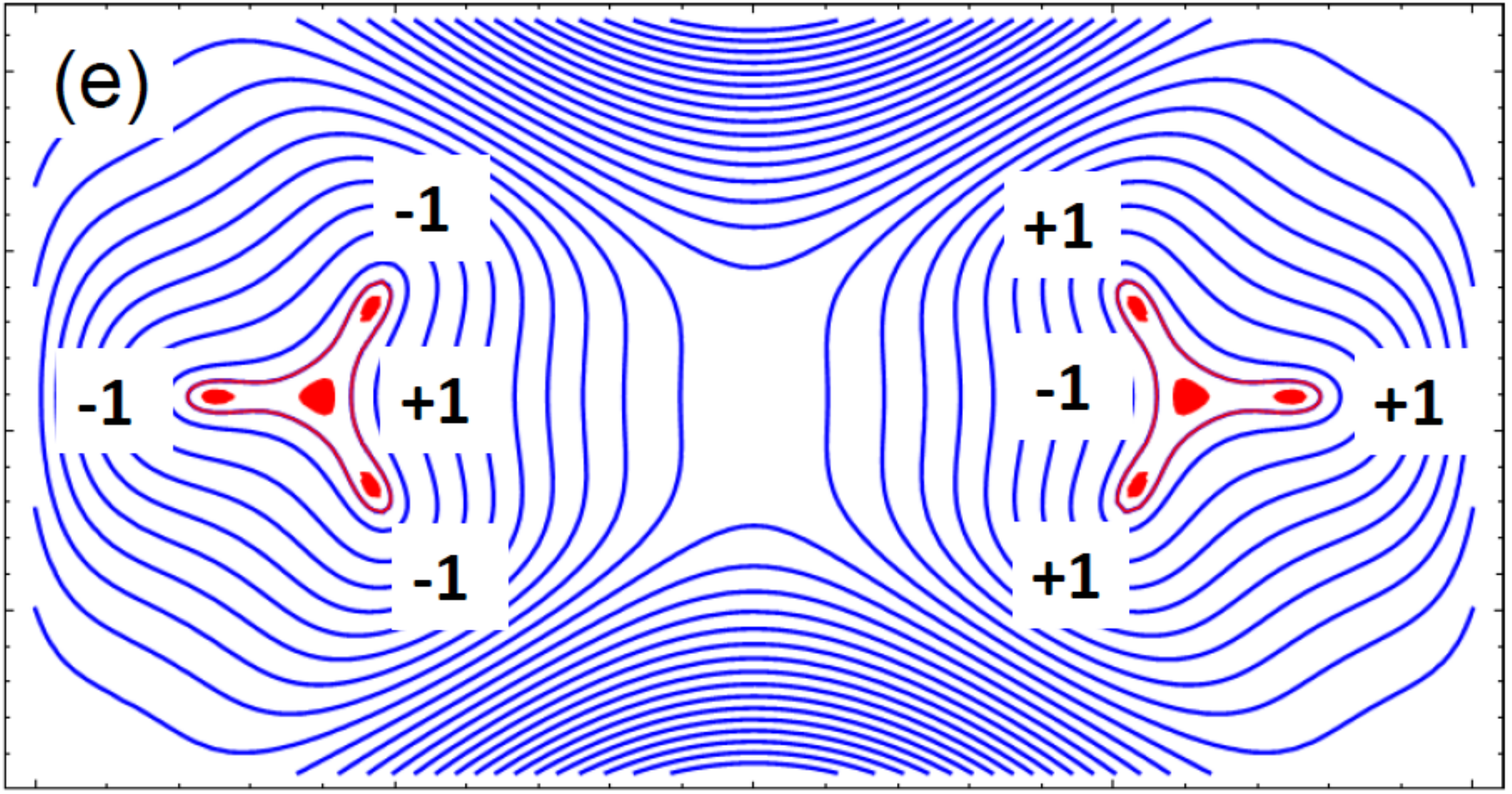}
\end{center}
\begin{center}
\includegraphics[width=8.5cm]{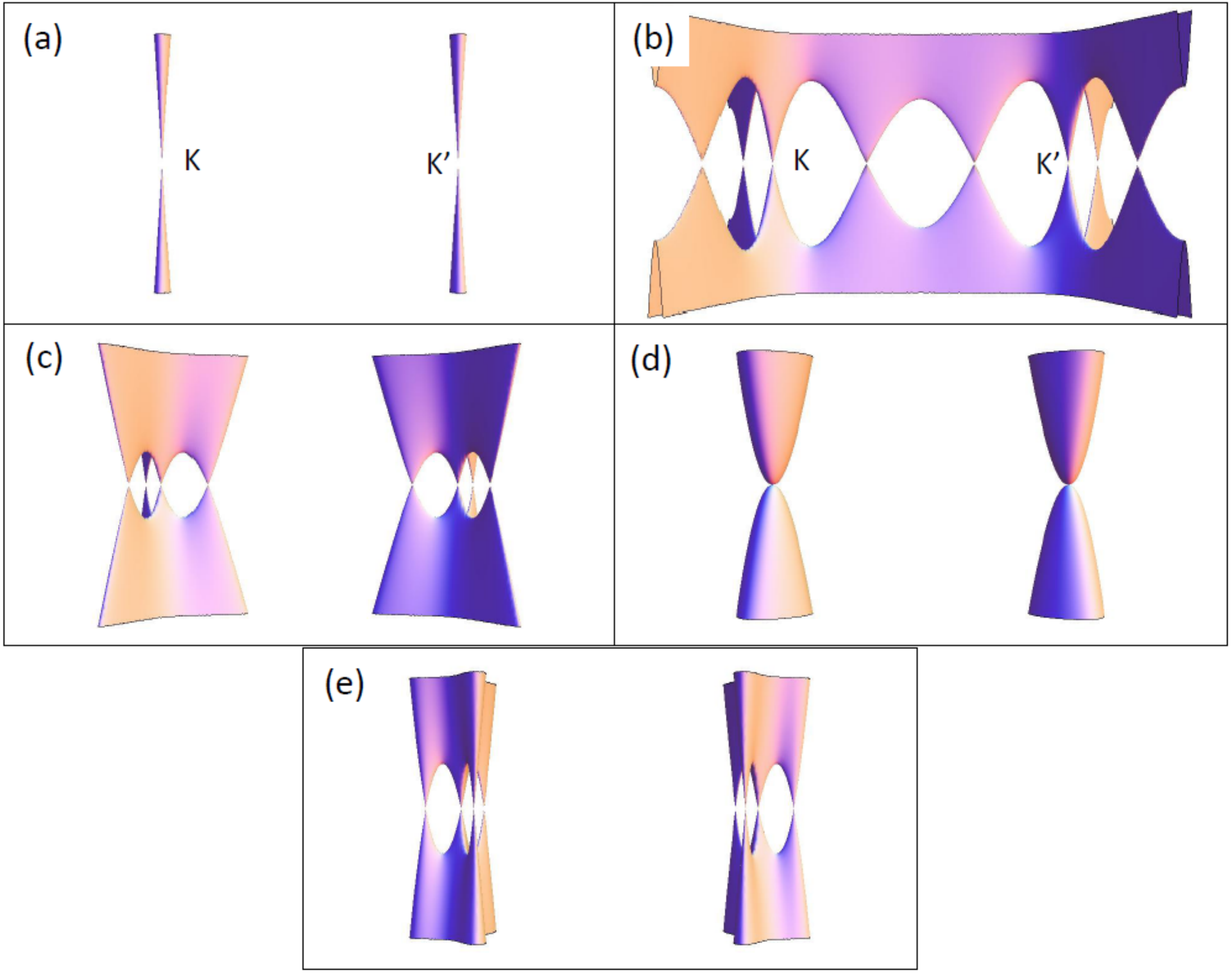}
\end{center}
\caption{Top: Iso-energy lines in the vicinity of the $\KK^{(')}$ and $\MM$ points in the $t-t_3$ model [Eqs. (\ref{GH}) and (\ref{f1f3})], for different values of the parameter $t_3$, (a) $t_3 = 0$, (b) $t_3 = 0.35 t$, (c) $t_3 = 0.40 t$, (d) $t_3 = 0.5 t$, , (e) $t_3 = 0.65 t$. The vicinity of the Dirac points is indicated in red, as well as their associated winding number (defined in section \ref{sect:winding}). Bottom: three-dimensional plot of the low energy spectrum for the same parameters.}
\label{fig:t3}
\end{figure}

In order to obtain additional pairs of Dirac points, the condition $f_{\D}=0$
necessarily implies more harmonics in the dispersion relation \cite{Montambaux:09,Sticlet}.
This can be achieved quite easily, at least in the framework of  a toy model, by adding a third-nearest-neighbors coupling $t_3$ in the tight-binding model of graphene. We do not
    consider  the coupling between second nearest neighbors  which, by coupling sites of the same sublattice, modifies the dispersion relation but does not affect the existence of Dirac points, as
long as the inversion symmetry is respected.\footnote{A coupling $t_2$ between second nearest neighbours dissymetrizes the spectrum. Interestingly, above a critical value $t_2=t/6$, there is a $1/\sqrt{energy}$ Van Hove singularity at the band edge \cite{Bellec:13b}} The Hamiltonian maintains the form (\ref{GH}), with the function
$f_{\k}$   given by \cite{remark1} (here $\beta=1$)

                 \be f_{\k} =t(\beta + e^{  i \k . \a_1} +e^{  i \k . \a_2}) + t_3 (e^{  i \k . (\a_1+\a_2)} + e^{  i \k . (\a_1-\a_2)} +e^{  i \k . (\a_2-\a_1)})\, \label{f1f3} \ee
In graphene, the $t_3$ term is small. However, it is of interest to imagine a larger value of this parameter because it has a quite interesting effect on the evolution of the spectrum, as has been theoretically considered in Refs. \cite{Bena:11,Montambaux:12}.
 When $t_3$ increases and reaches the critical value $t/3$, a new pair of Dirac points emerges from each of the three inequivalent ${\bf M}$ points in reciprocal space (see Fig. \ref{fig:t3}-a,b), following precisely the above universal scenario. As mentioned in Sec. \ref{sec:MMDP}, the annihilation
as well as the emergence of Dirac points occurs necessarily at TRIM that are precisely the ${\bf M}$
points at the border of the hexagonal BZ between the $K$ and $K'$ points.
Writing $\k = \M + \q$ , we recover the  Universal Hamiltonian in terms of the continuum wave vector
$\q$ in the vicinity of $t_3 =t/3$ (keeping the leading order terms)
      where the parameters $m^*, c, \Delta_*$ can be related to the original band parameters~: $\Delta_*= t - 3 t_3 $,  $c= 2 t$ and  $m^* =2 / t$. The parameter $\Delta_*$, when it becomes negative ($t_3 > t/3$), drives the emergence of a new pair of Dirac points at the $\MM$-point (Fig. \ref{fig:t3}-b,c). The distance between the new Dirac points is given by $2 q_D = 2\sqrt{- 2 m^* \Delta_*}= 4 \sqrt{3 t_3/t -1}$.
\medskip

We have thus added {\it three  pairs} of Dirac points, each pair emerging from one of the three $\MM$ points.
 When increasing further $t_3$, the new Dirac points approach the $\KK$ and $\KK'$ points, so that each initial Dirac point sitting at the $\KK^{(')}$ points is now surrounded by three Dirac points (with opposite charges, see Fig. \ref{fig:t3}.c.). These Dirac points merge at the critical value $t_3 =t/2$, and the spectrum becomes quadratic around $\KK^{(')}$   (Fig. \ref{fig:t3}-b,c) \cite{Bena:11,Montambaux:12}.

 Near $t_3 \simeq t/2$, the  Hamiltonian takes a new  form (keeping leading order terms) in the vicinity of the $\KK'$ point

 \be {\cal{H}}'(\q)= \left(
                \begin{array}{cc}
                 0 & \displaystyle  -{\bsq^2 \over 2 m^*}+ c \, \bsq^\dagger + \Delta \\
                \displaystyle      -{{\bsq^{\dagger}2} \over 2 m^*}+ c \, \bsq + \Delta^* & 0 \\
                \end{array}
              \right) \label{Hunivprime} \ee
where $\bsq=q_x + i q_y$. The Hamiltonian in  the vicinity of the $\KK$ point is obtained by the substitution  $\bsq \rightarrow - \bsq^\dagger$.
   Starting from (\ref{GH},\ref{f1f3}), we find $m^*= 4  / 9 t$ and $c=3  (t_3 - t/2) $.  When
$\Delta=0$, the low-energy Hamiltonian that of bilayer graphene \cite{MF}, and one obtains
moreover a parabolic band-contact point when $c=0$.
Within a tight-binding model, bilayer graphene is characterized essentially by three hopping integrals,
the coupling $\gamma_0$ between nearest neighbours in each layer (above named $t$), the coupling $\gamma_1$  between sites from different layers which
 are on top of each other,  and the coupling $\gamma_3$  between nearest-neighbour sites from different layers which do not face each other. Neglecting this third coupling, the quadratic low energy spectrum in each valley is described by a $2 \times 2$ Hamiltonian of the form (\ref{Hunivprime}) with $c = \Delta=0$, and a mass given by $m^*= 2 \gamma_1 / (9 \gamma_0^2)$.
For $\Delta=c=0$, the eigenstates of Hamiltonian (\ref{Hunivprime}) are given by the same expression
as those in Eq. (\ref{eq:b0states}), if one replaces $\phi_{\q}\rightarrow 2\phi_{\q}$. One thus
notices that the associated winding number around a parabolic band-contact point is $w=\pm 2$.

  The effect of the small $\gamma_3$ term is to induce a trigonal warping, so that the spectrum is no longer quadratic but consists of {\it  four} Dirac points. This is in agreement with the
additivity of topological charges discussed in Sec. \ref{sect:winding} -- indeed, the parabolic
band-contact point with $w=2$ is split into a central Dirac point with $w_{centr}=-1$ and
three additional Dirac points with $w_i=+1$, such that the sum gives again $w=w_{centr}+3w_i=2$.
This trigonal warping is described by Hamiltonian  (\ref{Hunivprime}) with $c= -3 \gamma_3/2$.
 The low-energy Hamiltonian for bilayers is thus equivalent to the Hamiltonian of the single layer with third nearest neighbours coupling, the correspondance being $t \leftrightarrow 2 \gamma_0^2/\gamma_1 + \gamma_3$ and $t_3 \leftrightarrow \gamma_0^2 / \gamma_1$ \cite{Montambaux:12}


\subsection{Manipulation of Dirac points in twisted bilayer: a second type of merging}
\label{sect:univ2}

Twisted bilayer graphene consists of two graphene layers that have a rotational mismatch with
respect to the conventional Bernal $AB$ stacking. In order to understand its low-energy spectrum,
consider for the moment two uncoupled layers that are rotated by a small angle $\theta$ with
respect to the $AB$-stacking reference. In this case, the two Dirac cones associated with the
two layers are separated in reciprocal space by a wave vector $\kappa$ that is a function of $\theta$.
Numerical calculations indicate that no gap is opened at the Fermi level when interlayer hopping
is taken into account \cite{numTBG}. However, the form of the
interlayer coupling fixes the relative winding number of one Dirac cone with respect to the other one
\cite{Degail:11}, and for small twist angles $\theta$ this coupling is continuously connected to
that in the ideal $AB$ case. The two Dirac points are therefore have the same winding number,
and, from a topological point of view, twisted bilayer at small angles is in the same class as
$AB$-stacked bilayer graphene.
Indeed, it can be described by the Hamiltonian
 \be {\cal{H}}_{++}(\q)= \frac{1}{2m^*}\left(
                \begin{array}{cc}
                 0 & \displaystyle  {\kappa^2 \over 4}-\bsq^2  \\
                \displaystyle      {\kappa^{*2} \over 4} -{\bsq^{\dagger2}}  & 0 \\
                \end{array}
              \right) \label{HunivprimeB} \ee
where the wave-vector shift is related to the gap parameter
$\Delta=\kappa^2/8m^* $ of Hamiltonian (\ref{Hunivprime}), with $c=0$ \cite{Degail:11,Degail:12}.
A finite value of $\kappa$ thus splits the quadratic dispersion relation into two cones separated by a saddle point (Fig. \ref{fig:bicones}-b).
This Hamiltonian describes the merging of two Dirac points with the {\it same } charge and has to be contrasted with the Hamiltonian (\ref{newH}) which describes the merging of Dirac points with {\it opposite} charge. In contrast to the latter case, discussed in Sec. \ref{sec:MMDP}, there is no
annihilation of the topological charges associated with the two Dirac points since one has a
topological transition from $w_1=+1$ and $w_2=+1$ (for $\kappa\neq 0$) to $w=+2$ at the merging.
The associated zero-energy Landau level in a magnetic field therefore remains two-fold degenerate from
a topological point of view (in addition to the usual four-fold spin-valley degeneracy) regardless of
the value of $\kappa$ \cite{Degail:11}, a scenario that has recently been verified experimentally
\cite{TBLexp}. We notice finally that the most general situation with $c \neq0$ and $\Delta \neq 0$ has been studied in Refs. \cite{Degail:12,coreans,Falko11,Ecrys}, in the framework of bilayer graphene,
where one layer is displaced by a constant vector with respect to the other one, with no twist
($\theta=0$).

\section{Conclusions}

In conclusion, we have discussed the basic properties of Dirac points that may occur  
in 2D crystalline systems, as well as their motion and merging. The physical systems that display
such Dirac points involve, apart from mono- and bi-layer graphene, graphene-like systems, such as
cold atoms in optical lattices, spatially
modulated semi-conductor heterostructures, quasi-2D organic crystals, microwave lattices,
molecular lattices, etc. Instead of an exhaustive discussion of all these artificial graphenes,
we have illustrated the theoretical aspects of Dirac-point motion in only some of them.
From the theoretical point of view, we have discussed some conditions for the emergence of
Dirac fermions in generic two-band models as well as the role of discrete symmetries, such as
time-reversal and inversion symmetry. Furthermore,
we have aimed at a classification of the different types of
Dirac point-merging, within a description of ``second-generation'' low-energy models and with
the help of a topological analysis in terms of winding numbers. These winding numbers, which
are revealed in the relative phase between the two components of the spinorial wave function,
may be interpreted as topological charges. Very much as electric charges, the winding numbers
are additive quantities, and their sum remains preserved in the different merging scenarios.
Whereas the merging of Dirac points with opposite winding number, such as in the case of time-reversal
symmetry related Dirac points, gives rise to a zero topological charge with the successive
annihilation of the Dirac points and the opening of a gap in the spectrum, the situation is
strikingly different in the case of Dirac-point merging with like topological charge. Indeed,
in this case, the band-contact points are preserved because of a non-zero winding number.
The set of parameters that give rise to a single (parabolic) band contact is singular in the sense
that a slight change in the parameters splits the parabolic contact point into two Dirac points,
as for example in the case of twisted bilayer graphene. The splitting into more than two Dirac
points is also possible, albeit with a sum of $\pm 2$ for the global topological charge, and occurs
for instance in bilayer graphene at very low energies, where trigonal warping becomes visible
-- in this case, one finds a central Dirac point with a winding number $w=-1$ surrounded by three
satellite Dirac points with $w=+1$ (for a total charge of $2$, here).

Apart from the stability of the band-contact points, the topological charges also allow us to
understand other physical quantities. They are revealed under the influence of a
magnetic field applied perpendicular to the 2D system. This field quantises the particles' energy
into discrete Landau levels that are highly degenerate from an orbital point of view. In addition
to this orbital degeneracy and the spin degree of freedom, one finds a topological degeneracy
of the zero-energy level that is precisely related to the topological charge. Whereas Dirac
points that are far apart in reciprocal space (as compared to the inverse magnetic length) provide
each a zero-energy Landau level, as stipulated by the low-energy Dirac-fermion model,
the situation becomes more complicated in the vicinity of the merging transitions. On the one
hand, the merging of time-reversal symmetry related Dirac points (with a zero total winding number)
destroys the (originally two-fold valley-degenerate) zero-energy level and splits it into two
seperate levels. On the other hand, the merging of two Dirac points with the same winding number
has no effect on the zero-energy level, which thus remains two-fold degenerate. The topological
charge (the total winding number) therefore indicates the number of topologically protected
zero-energy levels, as compared to the total number of possible zero-energy levels, which coincides
with the total number of Dirac points in the absence of a magnetic field.

\medskip

The work presented here results from several fruitful collaborations. First of all, we would like to
acknowledge the long-term in-house collaboration with our colleagues J.-N. Fuchs and F. Pi\'echon. Furthermore we would like to thank our students and postdocs R. de Gail, P. Delplace, P. Dietl, and L.-K. Lim, as well as our external collaborators M. Bellec, U. Kuhl, F. Mortessagne, F. Guinea, and 
A. H. Castro Neto.

\end{document}